\documentclass[12pt]{article} 
\usepackage{a41}
\usepackage{epsfig}
\usepackage{cite}
\usepackage{amsmath}
\usepackage{amssymb}
\usepackage{float}
\usepackage{verbatim}
\usepackage{amsxtra}
\usepackage{afterpage}

\newcommand{\SigmaP}{\textsf{Sigma}}

\newcommand{\D}{\displaystyle} 
\newcommand{\N}{\nonumber} 
\newcommand{\ep}{\varepsilon}
\newcommand\eph{\frac{\varepsilon}{2}}
\newcommand{\ph}[2]{\left(#1\right)_{#2}}  
\newcommand{\Gam}[2]{\Gamma\left[\begin{array}{c}#1\\#2\end{array}\right]} 

\newcommand{\bea}{\begin{eqnarray}}
\newcommand{\bq}{\begin{equation}}
\newcommand{\eea}{\end{eqnarray}}
\newcommand{\eq}{\end{equation}}

\newcommand{\gsim}{\raisebox{-0.07cm   }
{$\, \stackrel{>}{{\scriptstyle\sim}}\, $}}

\newcommand\LL{\mbox{\rm L}}
\newcommand\HH{\mbox{\rm H}}

\newcommand\be{\begin{eqnarray}}
\newcommand\ee{\end{eqnarray}}

\newcommand\Mvec{\,\mbox{\bf M}}
\begin{document}
\noindent
\sloppy
\thispagestyle{empty}
\begin{flushleft}
DESY 12-056 \hfill 
\\
DO-TH-12/17 \\
TTK-12-26\\
SFB/CPP-12-37 \\ 
LPN 12-055 \\
June 2012
\end{flushleft}
%
\vspace*{\fill}
\hspace{-3mm}
{\begin{center}
{\Large\bfseries Massive 3-loop Ladder Diagrams for}

\vspace*{2mm}
{\Large\bfseries Quarkonic Local Operator Matrix Elements}

\end{center}
}

\begin{center}
\vspace{2cm}
\large
Jakob Ablinger$^a$, Johannes Bl\"umlein$^b$, Alexander Hasselhuhn$^b$,\\
Sebastian Klein$^c$, Carsten Schneider$^a$, and Fabian Wi\ss{}brock$^b$
\vspace{5mm}
\\

\vspace{5mm}
\normalsize {\itshape $^a$~Research
Institute for Symbolic Computation (RISC),\\ Johannes Kepler
University, Altenbergerstra\ss{}e 69, A-4040 Linz, Austria}
\\ 

\vspace{5mm}
\normalsize
{\itshape $^b$~Deutsches Elektronen--Synchrotron, DESY,\\
Platanenallee 6, D--15738 Zeuthen, Germany}
\\

\vspace{5mm}
\normalsize {\itshape $^c$~Research
Institut f\"ur Theoretische Physik E, RWTH Aachen University, D--52056 Aachen,
Germany} \\ \vspace{2em}
\end{center}

\vspace*{\fill} %
\begin{abstract}
\noindent
3-loop diagrams of the ladder-type, which emerge for local quarkonic twist-2 
operator matrix elements, are computed directly for general values of the Mellin 
variable $N$ using Appell-function representations and applying modern summation 
technologies provided by the package {\sf Sigma} and the method of hyperlogarithms. 
In some of the diagrams generalized harmonic sums with $\xi \in \{1,1/2,2\}$ 
emerge beyond the usual nested harmonic sums. As the asymptotic representation
of the corresponding integrals shows, the generalized sums conspire giving
well behaved expressions for large values of $N$. These diagrams contribute to the 
3-loop heavy flavor Wilson coefficients of the structure functions in deep-inelastic 
scattering in the region $Q^2 \gg m^2$.
\end{abstract}

\vspace*{\fill} 

\newpage
\section{Introduction}
\label{sec:1}
\renewcommand{\theequation}{\thesection.\arabic{equation}}
\setcounter{equation}{0} 

\vspace{1mm}
\noindent
The Wilson coefficients for the twist-2 heavy flavor contributions to the 
unpolarized structure functions in deeply inelastic scattering are known in
leading \cite{LO} and next-to-leading order \cite{NLO}~\footnote{For a fast 
and precise numerical Mellin-space implementation see \cite{AB}.}. In the 
latter case, the result was obtained in semi-analytic form. It has been shown
in \cite{BUZA1} that in the region $Q^2 \gg m^2$ one can obtain analytic 
representations due to a factorization relation of the heavy flavor Wilson coefficients 
being valid for all contributions but the power corrections $\propto (m^2/Q^2)^k,~~k \geq 1$. 
In this representation the heavy flavor Wilson coefficients are given as convolutions 
between {\sf universal} massive operator matrix elements (OMEs) and the massless Wilson 
coefficients. The final results in case of the previous calculations were most compactly 
expressed in Mellin space in terms of harmonic sums \cite{HSUM,SUMMER}, which allows further 
simplifications concerning basis representations, cf.~\cite{HBAS1,HBAS2}. The 2-loop 
corrections 
in the neutral current case have been calculated in 
\cite{BUZA1,Bierenbaum:2007qe,BBK3,BBK4,Buza:1996xr,Bierenbaum:2007pn} and for the charged 
current case up to NLO  for $Q^2 \gg m^2$ in \cite{CC}. For the structure function $F_L(x,Q^2)$ 
the general result in the asymptotic case was obtained at 3-loop order in Ref.~\cite{BFNK}. 
Unlike the case for the structure function $F_2(x,Q^2)$, where the approximation works at 
the per cent level for $Q^2 \gsim 10~m^2$, much larger scales are needed in the former 
case. 
Recently, complete results have been obtained for a series of Mellin moments 
$N =2 \ldots 10 (14)$ for the different heavy flavor Wilson coefficients 
contributing to the structure function $F_2(x,Q^2)$ in Ref.~\cite{BBK2}\footnote{
The corresponding results in case of transversity were given in \cite{TRANSV}.},
including all operator matrix elements needed to establish the variable flavor
scheme \cite{BUZA2,BBK3}. Here all logarithmic contributions  $\ln^k(Q^2/m^2),~k = 1, 2, 3$, 
are complete for general values of $N$, \cite{Bierenbaum:2010jp}, 
referring to the anomalous dimensions and Wilson coefficients known in the literature 
\cite{MVV1,WIL2,MVV2} as well as linear terms in the dimensional expansion parameter 
$\varepsilon = D - 4$ at 2-loop order, \cite{BBK4}. The constant term has been calculated 
for the color factors $T_f^2 n_f C_{A,F}$ at NNLO for the quarkonic and gluonic massive 
OMEs in \cite{Ablinger:2010ty,Blumlein:2012vq} and first contributions to $T_f^2  
C_{A,F}$ were given in \cite{Ablinger:2011pb}.

In the present paper we compute  diagrams of the ladder topology with up to six massive 
propagators contributing to the massive 3-loop operator matrix elements for 
general values of $N$ as another step towards the general $N$ result. The graphs 
are computed directly, i.e. without reference to the integration-by-parts 
method~\cite{IBP}~\footnote{We also refrain from using Mellin-Barnes integral 
representations \cite{MB}, which would lead to complications instead of 
simplifications in the present case.}. We seek for closed representations in 
$D$-dimensions referring to suitable higher transcendental functions. At the 
2-loop level investigated formerly in Ref.~\cite{BBK5} the corresponding class 
is formed by the generalized hypergeometric functions $_pF_q$ \cite{GHYP,Slater}, 
while in the present case Appell-functions emerge \cite{APPELL} as 
characteristics of the topology being dealt with. The Feynman diagrams are
represented by multiple nested sums of products of higher transcendental 
functions which are easily expanded in $\varepsilon$. These sums are partly 
due to the local twist-2 operator insertions, cf.~Ref.~\cite{BBK2}, Appendix~8.1, 
and partly due to binomial expansions of the Feynman parameter integrals. Finally, 
these sums are evaluated using modern summation technologies  which are 
made available by the code {\sf Sigma} \cite{SIGMA} and are expressed in sums being 
transcendental to each other, and multiple zeta values~\cite{MZV} or related 
constants, which constitute the values at $N \rightarrow \infty$ of the former 
ones. We also use the method of hyperlogarithms \cite{BROWN}, having been extended to 
the case of local operator insertions at general values of $N$.
For low enough fixed values of $N$ one may compare the results
with those obtained by computing the corresponding diagrams using {\sf MATAD} 
\cite{MATAD}. The method presented in the present paper is applicable for a 
wider range of processes.

The paper is organized as follows. In Section~2 the basic diagram, without 
operator insertion, is analyzed and a sample calculation is presented. The 
main different cases are evaluated in Section~3. We also provide a series 
of fixed moments used to compare the results of the present calculation with 
those obtained using {\sf MATAD}. We discuss the origin of $S$-sums 
\cite{MUW,ABS1}\footnote{Another generalization of harmonic sums and polylogarithms
are those generated by cyclotomic polynomials, cf. \cite{Ablinger:2011te}.} 
generated  of argument $\xi \in \{1,1/2,2\}$, which emerge in the final 
result of some of the diagrams 
besides the usual harmonic sums. Knowing the result for the different 
diagrams we also apply the method developed in Ref.~\cite{LINEQ} in Section~4 to 
establish and solve difference equations from fixed moments and determine the minimal
amount of moments needed to reconstruct the corresponding $N$-dependent functions
completely. Section~5 contains the conclusions.
\section{Basic Formalism}
\label{sec:2}
\renewcommand{\theequation}{\thesection.\arabic{equation}}
\setcounter{equation}{0} 

\vspace{1mm}
\noindent
We consider diagrams of the type shown in Figure~1 and those related to them, cf. also \cite{SKPHD}. 
    \begin{figure}[H]
     \centering
     \includegraphics[angle=0, height=3.3cm]{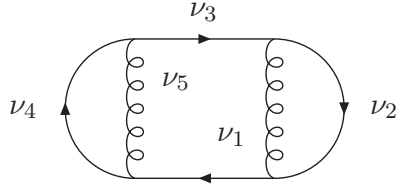}
     \caption{A $3$--loop tadpole ladder graph.}
     \label{3Lb}
    \end{figure} 
\noindent
The scalar $D$--dimensional integral corresponding to the diagram in Figure~1 for 
arbitrary exponents $\nu_i$ of the propagators reads~:
\begin{eqnarray}
T_1&=&\iiint \frac{d^Dqd^Dkd^Dl}{(2\pi)^{3D}} 
                  \frac{i (-1)^{\nu_{12345}} (m^2)^{\nu_{12345}-3D/2}
                        (4\pi)^{3D/2}}
                        {
                         (k^2)^{\nu_1}((k-l)^2-m^2)^{\nu_2}
                         (l^2-m^2)^{\nu_3}
                         ((q-l)^2-m^2)^{\nu_4}(q^2)^{\nu_5}
                        }~,
\end{eqnarray}
where $\nu_{i_1 i_2 ... i_l} = \sum_{k=1}^l \nu_i$ and suitable normalization factors have been attached for convenience. 
The loop--momenta are integrated in the order $k,q,l$. The following
Feynman--parameter representation is obtained~:
\begin{eqnarray}
T_1&=&
           \Gam{\nu_{12345}-6-3\ep/2}{\nu_1,\nu_2,\nu_3,\nu_4,\nu_5} 
\N\\ &&
           \int_0^1~dw_1\ldots \int_0^1~dw_4
           \frac{\theta(1-w_1-w_2) w_1^{-3-\ep/2+\nu_{12}}w_2^{-3-\ep/2+\nu_{45}}(1-w_1-w_2)^{\nu_3-1}}
           {\D{\left(1+w_1\frac{w_3}{1-w_3}+w_2\frac{w_4}{1-w_4}\right)^{\nu_{12345}-6-3\ep/2}}}
\N\\ \N\\&&
       \times
                  w_3^{1+\ep/2-\nu_1}(1-w_3)^{1+\ep/2-\nu_{2}}
                  w_4^{1+\ep/2-\nu_5}(1-w_4)^{1+\ep/2-\nu_{4}}~,
                  \label{bubble3}
\end{eqnarray}
with the short--hand notation
\begin{eqnarray}
    \Gam{a_1, ..., a_n}{b_1, ..., b_m}
    = \prod_{k=1}^n \prod_{l=1}^m \Gamma(a_k) \Gamma^{-1}(b_l)~.    
\end{eqnarray}
In order to perform the $\{w_1,~w_2\}$ integration, one considers
\begin{eqnarray}
\label{EQ1}
I&=&\int_0^1dw_1\int_0^1dw_2~\theta(1-w_1-w_2)w_1^{b-1}w_2^{b'-1}
                          (1-w_1-w_2)^{c-b-b'-1}(1-w_1x-w_2y)^{-a},
\N\\
\end{eqnarray}
with the parameters $a,b,b',c$ such that this integral is convergent. 
Eq.~(\ref{EQ1}) can then be expressed in terms of the Appell-function $F_1$
using the relation, \cite{Slater}~\footnote{Note that Eq.~(8.2.2) 
of Ref.~\cite{Slater} contains typographical errors.},
\begin{eqnarray}
     I&=&\Gam{b,b',c-b-b'}{c}
         \sum_{m,n=0}^{\infty}
         \frac{\ph{a}{m+n}\ph{b}{n}\ph{b'}{m}}
           {\ph{1}{m}\ph{1}{n}\ph{c}{m+n}}
         x^ny^m
\N\\
     &=& \Gam{b,b',c-b-b'}{c}
          F_1\Bigl[a;b,b';c;x,y\Bigr].   
     \label{F1def}
\end{eqnarray}
Here the parameters $x,y$ correspond to $w_3/(w_3-1)$ and $w_4/(w_4-1)$ 
in Eq.~(\ref{bubble3}), respectively. To obtain a series-representation
of the integral 
the analytic continuation of $F_1$, \cite{Slater},
\begin{eqnarray}
     F_1\left[a;b,b';c;\frac{x}{x-1},\frac{y}{y-1}\right] =
         (1-x)^b(1-y)^{b'}F_1[c-a;b,b';c;x,y]  
\label{F1ancont}
\end{eqnarray}
has to be carried out. One obtains the infinite double sum
\begin{eqnarray}
T_1&=&
           \Gam{-2-\ep/2+\nu_{12},-2-\ep/2+\nu_{45},-6-3\ep/2+\nu_{12345}}{\nu_2,\nu_4,-4-\ep+\nu_{12345}}
\N\\ && \times
           \sum_{m,n=0}^{\infty}
           \Gam{2+m+\ep/2-\nu_1,2+n+\ep/2-\nu_5}{1+m,1+n,2+m+\ep/2,2+n+\ep/2}
\N\\ && \times
           \frac{\ph{2+\ep/2}{n+m}\ph{-2-\ep/2+\nu_{12}}
{m}\ph{-2-\ep/2+\nu_{45}}{n}}{\ph{-4-\ep+\nu_{12345}}{n+m}}~. 
\label{bubble3Res}
\end{eqnarray} 
Here $\ph{a}{b}$ is Pochhammer's symbol defined by
\begin{eqnarray}
    \ph{a}{b} = \frac{\Gamma(a+b)}{\Gamma(a)}~.
\end{eqnarray} 
Eq.~(\ref{bubble3Res}) is symmetric w.r.t. exchanges of the indices 
$\{\nu_1,~\nu_2\}~\leftrightarrow~\{\nu_4,\nu_5\}$. For any values of $\nu_i$ of the 
type $\nu_i=a_i+b_i\ep$, with $a_i\in~{\mathbb N},~b_i\in~{\mathbb C}$, the Laurent--series 
in $\ep$ can be calculated straight-forwardly using e.g. {\sf summer}, 
\cite{SUMMER} or {\sf Sigma} \cite{SIGMA}. 
We have  checked (\ref{bubble3Res}) for various values of the $\nu_i$ using ${\sf MATAD}$~\cite{MATAD}.

\begin{figure}[H]
     \centering
     \includegraphics[angle=0, height=3.3cm]{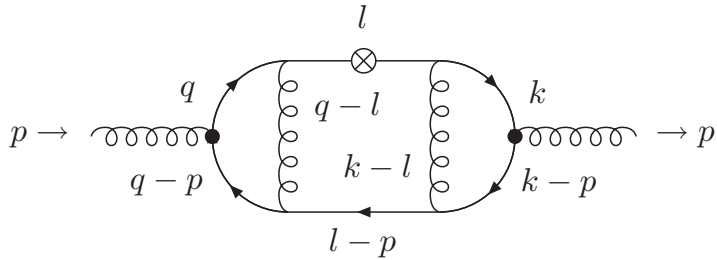}
     \caption{The $3$--loop ladder graph containing a central local operator insertion.}
     \label{3Lc}
    \end{figure}

We now consider the diagram shown in Figure~\ref{3Lc}, which contributes to the massive operator matrix 
element $A_{Qg}^{(3)}$, cf.~Ref.~\cite{BBK2}, and derives from the diagram in Figure~1, 
by adding the local operator 
insertion $\otimes$, see \cite{BBK2} and Appendix~A, and an external momentum flow $p$ with $p^2 = 0$. We consider first 
the case where all exponents of the propagators are equal to one.
    \noindent
Here and in the following we separate a common  pre-factor 
\begin{eqnarray}
  I_{1a} \equiv \frac{i(\Delta.p)^Na_s^3S_\ep^3}{(m^2)^{2-3\eph}}\hat{I}_{1a}.
\end{eqnarray} 
Again the momentum integrals are performed in the order $q,k,l$ through which
\begin{eqnarray}
   \hat{I}_{1a}
   &=& - \exp\left(-3\eph\gamma_E\right) 
      \Gamma(2-3\ep/2) \prod_{i=1}^7 \int_0^1 dw_i~
      \frac{\theta(1-w_1-w_2) w_1^{-\ep/2} w_2^{-\ep/2} (1-w_1-w_2)}
        {\D{\left(1+w_1\frac{1-w_3}{w_3}+w_2\frac{1-w_4}{w_4}\right)^{2-3\ep/2}}} 
\N \\&&
      \times
      w_3^{-1+\ep/2}
      (1-w_3)^{\ep/2}
      w_4^{-1+\ep/2}
      (1-w_4)^{\ep/2}
      (1-w_5w_1-w_6w_2-(1-w_1-w_2)w_7)^N \N\\ 
\label{IL1}
\end{eqnarray}
is obtained. The spherical factor $S_\ep$ is given by
\begin{eqnarray}
    S_\ep = \exp\left[\eph\left(\gamma_E - \ln(4\pi)\right)\right]~,
\end{eqnarray}
and $\gamma_E$ the Euler--Mascheroni constant. In the $\overline{\rm MS}$ scheme
this factor is later identified by $S_\ep \equiv 1$. It absorbs the universal dependence
of $D$-dimensional integrals on $\ln(4\pi)$ and $\gamma_E$. As in the $2$--loop case, 
\cite{BBK5}, one observes that the integral--kernel given by the corresponding massive 
tadpole--integral (\ref{bubble3}) 
is multiplied by a polynomial containing various integration parameters to the power $N$. 
The same holds 
for the remaining $3$--loop diagrams. Hence, if a general sum representation for the 
corresponding tadpole--integrals is known and one knows how to evaluate the 
corresponding sums, the $3$--loop massive OMEs can be calculated directly. 
However,
the presence of the polynomial to the power $N$, which may also 
involve a finite sum, cf. the Feynman rules given in  Appendix~8.1. of Ref.~\cite{BBK2}, complicates the 
calculation 
further, in some cases even in a {\it very essential} way. 

To perform the remaining integrals for Diagram $I_{1a}$ we split the expression into several finite sums which have
a similar structure as $T_1$.  One obtains
\begin{eqnarray} 
   \hat{I}_{1a}
   &=& 
     \frac{\exp\left(-3\eph\gamma_E\right)\Gamma(2-3\ep/2)}
        {(N+1)(N+2)(N+3)}\sum_{m,n=0}^{\infty}
      \Biggl\{
\N\\ && 
\hspace{5mm}
      \sum_{t=1}^{N+2}
      \binom{3+N}{t}
      \frac{\ph{t-\ep/2}{m}\ph{2+N+\ep/2}{n+m}\ph{3-t+N-\ep/2}{n}}
        {\ph{4+N-\ep}{n+m}}
\N \\ && 
\hspace{10mm} 
      \times
      \Gam{t,t-\ep/2,1+m+\ep/2,1+n+\ep/2,3-t+N,3-t+N-\ep/2}
        {4+N-\ep,1+m,1+n,1+t+m+\ep/2,4-t+n+N+\ep/2}
\N\\ && 
\hspace{1mm}
     -\sum_{s=1}^{N+3}
      \sum_{r=1}^{s-1}
      \binom{s}{r}
      \binom{3+N}{s}(-1)^s
      \frac{\ph{r-\ep/2}{m}\ph{-1+s+\ep/2}{n+m}\ph{s-r-\ep/2}{n}}
        {\ph{1+s-\ep}{n+m}}
\N\\ && 
\hspace{10mm} 
      \times
      \Gam{r,r-\ep/2,s-r,1+m+\ep/2,1+n+\ep/2,s-r-\ep/2}{1+m,1+n,1+r+m+\ep/2,1+s-r+n+\ep/2,1+s-\ep}
      \Biggr\}~. 
\label{IL2}
\N\\
\end{eqnarray}
After expanding in $\ep$, the summation can be performed using \SigmaP~and 
the summation techniques having been explained in \cite{Bierenbaum:2007qe,BBK4,Blumlein:2012hg} before.
 
The results of the integrals being dealt with in the present paper can be expressed
in terms of  harmonic sums $S_{\vec{a}}(N)$ \cite{HSUM,SUMMER} and their 
generalizations $S_{\vec{a}}(\vec{\xi};N)$ \cite{MUW,ABS1}. They are defined by~: 
\begin{eqnarray}
S_{b,\vec{a}}(N) &=& \sum_{k=1}^N \frac{{{\rm sign}(b)}^k}{k^{|b|}} 
S_{\vec{a}}(k),~~~~S_\emptyset(k) = 1~,
\end{eqnarray}
and
\begin{eqnarray}
S_{b,\vec{a}}(\eta,\vec{\xi};N) &=& \sum_{k=1}^N \frac{\eta^k}{k^{b}} S_{\vec{a}}(\vec{\xi};k),~~S_\emptyset = 
1,~~~~\eta, \xi \in \mathbb{R}, b,a_i \in \mathbb{N} \backslash \{0\}~. 
\end{eqnarray}
In the following we use the short-hand notation for these sums $S_{\vec{a}}(N) \equiv 
S_{\vec{a}}$, 
$S_{\vec{a}}(\vec{\xi};N) \equiv S_{\vec{a}}(\vec{\xi})$. 

Moreover, we calculate the diagrams for all integer values of $N$, for which they are defined and do not refer either 
to even or odd moments, as required in subsequent physical applications due to the presence of the respective current
crossing relations, cf.~\cite{CROSS}. 

The threefold and fourfold sums in $\hat{I}_{1a}$ yield~:
    \begin{eqnarray}
      \hat{I}_{1a}&=&
       -\frac{4(N+1){S_1}+4}{(N+1)^2(N+2)}{\zeta_3}
       +\frac{2{S_{2,1,1}}}{(N+2)(N+3)}
       +\frac{1}{(N+1)(N+2)(N+3)}\Biggl\{                        
\N\\ &&
                        -2(3N+5){S_{3,1}}
                        -\frac{S_1^4}{4}
                        +\frac{4(N+1){S_1}-4N}{N+1}{S_{2,1}}
                        +2\left[
                           (2N+3){S_1}
                          +\frac{5N+6}{N+1}
                          \right]{S_3}
\N\\ &&
                        +\frac{9+4N}{4}{S_2^2}
                        +\left[
                           2\frac{7N+11}{(N+1)(N+2)}
                          +\frac{5N}{N+1}{S_1}
                          -\frac{5}{2}{S_1^2}
                         \right]{S_2}
                        +\frac{2(3N+5) S_1^2}{(N+1)(N+2)}
\N\\ &&
                        +\frac{N}{N+1}{S_1^3}
                        +\frac{4(2N+3){S_1}}{(N+1)^2(N+2)}
                        -\frac{(2N+3){S_4}}{2}
                        +8\frac{2N+3}{(N+1)^3(N+2)}
                               \Biggr\}
\N\\ &&                +O(\ep)~,
    \end{eqnarray}
which agrees with the fixed moments $N=1\ldots 10$ obtained using ${\sf MATAD}$~\cite{MATAD}.
For a direct reference we give a series of moments in Tables~1 and 2 below. 
    
The propagator carrying the operator insertion in Figure~2 also emerges to the second power.
 The result for the corresponding integral is
\begin{eqnarray}
I_{1b} &\equiv&  \frac{i(\Delta.p)^Na_s^3S_\ep^3}{(m^2)^{3-3\eph}}\hat{I}_{1b}~,
\\
\hat{I}_{1b} &=&  
   \frac{ \exp\left(-3\eph\gamma_E\right)}
        {(N+1) (N+2) (N+3)} 
   \Gamma\left(3 - \frac{3}{2} \ep\right)
   \Biggl\{ 
  -\sum_{m=0}^{\infty} 
   \sum_{n=0}^{\infty}  
   \sum_{l=1}^{N+2} \binom{N+3}{l} 
\N\\ && 
\times 
   B\left(l,m+1+\eph\right) 
   B\left(N+3-l,n+1+\eph\right)   
\N\\ && 
\times                
   \Gam{N+2+\eph+m+n}{m+1, n+1, N+2 +\eph}
   \frac{B\left(l+m-\eph, N+3-l+n -\eph\right)}
     {(N+4+m+n-\ep) (N+3+m+n-\ep)} 
\N\\&&
  +\frac{1}{N+4} \Biggl[
   \sum_{m=1}^{\infty} 
   \sum_{n=1}^{\infty}  
   \sum_{l=1}^{N+4} \binom{N+4}{l} 
   \sum_{j=1}^{N+4-l}\binom{N+4-l}{j} (-1)^{j+l}
   B\left(j, m+1 +\eph\right)
\N\\ && 
\times                
   B\left(l, n+1 +\eph\right)
   \Gam{j+l-2+m+n+ \eph}{m+1, n+1, j+l-2 +\eph} 
   \frac{B\left(j+m-\eph, l+n-\eph\right)}
     {j+l+m+n-\ep} 
\N
\end{eqnarray}
\begin{eqnarray}
&&
  +\sum_{m=1}^{\infty} 
   \sum_{l=1}^{N+4} \binom{N+4}{l} 
   \sum_{j=1}^{N+4-l}\binom{N+4-l}{j} (-1)^{j+l}
   B\left(j, m+1 +\eph\right)
   B\left(l, 1 +\eph\right)
\N\\ && 
\times                
   \Gam{j+l-2+m+ \eph}{m+1, j+l-2 +\eph} 
   \frac{B\left(j+m-\eph, l-\eph\right)}
     {j+l+m-\ep} 
\N\\
&&
   +\sum_{n=1}^{\infty}
   \sum_{l=1}^{N+4} \binom{N+4}{l} 
   \sum_{j=1}^{N+4-l}\binom{N+4-l}{j} (-1)^{j+l}
   B\left(j, 1 +\eph\right)
   B\left(l, n+1 +\eph\right)
\N
\\ 
&& 
\times                
   \Gam{j+l-2+n+ \eph}{n+1, j+l-2 +\eph} 
   \frac{B\left(j-\eph, l+n-\eph\right)}
     {j+l+n-\ep} 
\N\\ &&
  +\sum_{l=1}^{N+4} \binom{N+4}{l} 
   \sum_{j=1}^{N+4-l}\binom{N+4-l}{j} (-1)^{j+l}
   B\left(j, 1 +\eph\right)
   B\left(l, 1 +\eph\right)
   \frac{B\left(j-\eph, l-\eph\right)}
        {j+l-\ep} 
\N\\
&&
  -\sum_{m=0}^{\infty}
   \sum_{n=0}^{\infty}
   \sum_{l=1}^{N+3} \binom{N+4}{l} 
   B\left(l,m+1+\eph\right)
   B\left(N+4-l, n+1+\eph\right) 
\N\\ && 
\times
   \Gam{N+2+m+n+\eph}{m+1, n+1, N+2 +\eph} 
   \frac{B\left(l+m-\eph, N+4-l+n-\eph\right)}
     {N+4+m+n-\ep} \Biggr]\Biggr\} 
\\
&=& 
   \frac{1}{(N+1) (N+2) (N+3) (N+4)} \Biggl\{
   \frac{1}{2} S_1^4 
  -\frac{3N+1}{N+1} S_1^3
\N\\&&
  -\frac{N^5+8 N^4+45 N^3+154 N^2+234 N+122}{(N+1)^2 (N+2) (N+3) } S_1^2
  +\frac{4 \left(5 N^3+22 N^2+23 N+3\right)}{(N+1)^2 (N+2) (N+3)} S_1
\N\\ & &
  -\frac{1}{2} \left(2 N^2+14 N+21\right) S_2^2
  -\frac{2 \left(6 N^5+46 N^4+170 N^3+411 N^2+575 N+324\right)}
     {(N+1)^3 (N+2)^2 (N+3)}
\N\\& &
  + 4 (N+3) (N+4) \left[
      S_1 
     +\frac{1}{N+1} 
   \right] \zeta_3
  +\Biggl[
    5 S_1^2
   -\frac{5 (3 N+1)}{(N+1)} S_1
\N
\end{eqnarray}
\begin{eqnarray}
&&
   -\frac{3 N^5+28 N^4+151 N^3+458 N^2+638 N+318}
      {(N+1)^2 (N+2) (N+3) }
   \Biggr] S_2
  +\Biggl[
    -\frac{2 \left(2 N^3+16 N^2+51 N+43\right)}{(N+1)}
\N\\ & &
  -4\left(N^2+7 N+9\right) S_1 
   \Biggr] S_3
  +\left(N^2+7 N+9\right)  S_4
  +\Biggl[
     \frac{2 \left(N^3+7 N^2+20 N+10\right)}{(N+1)}
  -8 S_1 
   \Biggr] S_{2,1}
\N\\ &&
  +2 \left(3 N^2+21 N+28\right) S_{3,1}
  -2 \left(N^2+7 N+8\right) S_{2,1,1} 
  \Biggr\}+ O(\ep)~,
\end{eqnarray}
and follows carrying out similar steps as in case of integral $I_{1a}$.
Here, 
\begin{eqnarray}
B(a,b) = \Gam{a,b}{a+b}
\end{eqnarray}
denotes Euler's Beta-function. The structure of the results of the integrals
$I_{1a}$ and $I_{1b}$ are very similar w.r.t. the harmonic sums. However, the polynomial
structure in $N$ becomes more involved for $I_{1b}$.

\section{Different Operator Insertions and Fermion Flows}
\label{sec:3}
\renewcommand{\theequation}{\thesection.\arabic{equation}}
\setcounter{equation}{0} 

\vspace{1mm}
\noindent
In the following we perform the calculation of ladder-type diagrams of
different complexity, which is both due to the number of massive lines and the
corresponding local operator insertions. The examples cover the spectrum of
possibilities. First we compute diagrams with six massive lines and a second
set of examples deals with diagrams which contain three massive lines. The
Feynman rules used for the calculation are given in Appendix~\ref{app:FR}.
\subsection{Diagrams with Six Fermion Propagators}
\label{sec:3.1}

\begin{figure}[htbp]
 \centering
 \parbox[c]{\textwidth}{
  \centering\epsfig{figure=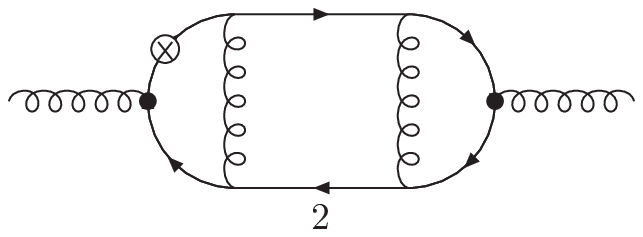,width=0.27\linewidth} \hspace{3mm}
  \centering\epsfig{figure=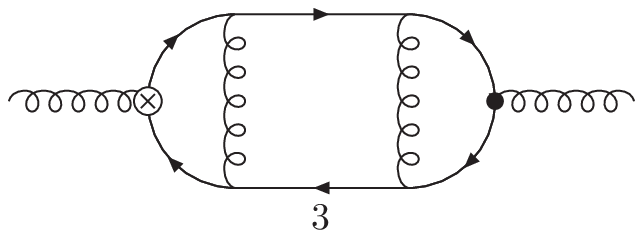,width=0.27\linewidth} \hspace{3mm}
  \centering\epsfig{figure=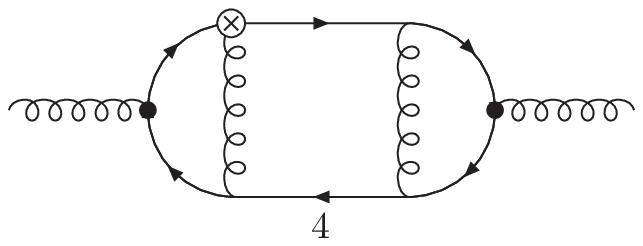,width=0.27\linewidth}
 }
 \caption{Diagrams with 6 fermion propagators}
 \label{fig:Diag6fp}
\end{figure}

\vspace{1mm}
\noindent
We consider the diagrams in Figure~\ref{fig:Diag6fp}.
In Table~1
we summarize a series of Mellin moments for the diagrams calculated 
using the code {\sf MATAD}~\cite{MATAD} for comparison to the general-$N$ results.

Diagram~$2a$ can be represented in terms of up to sixfold sums.
\begin{eqnarray}
 I_{2a} &\equiv& \frac{i(\Delta.p)^Na_s^3S_\ep^3}{(m^2)^{2-3\eph}}\hat{I}_{2a},\\
 \hat{I}_{2a}
 &=&   
    \exp\left(-3\eph\gamma_E\right)
    \Gamma\left(2 - \frac{3}{2}\ep\right) 
    \frac{1}{(N+1) (N+2)}
    \sum_{m=0}^{\infty}
    \sum_{n=0}^{\infty}
    \sum_{l=2}^{N+2}
    \binom{N+2}{l}
    \sum_{j=2}^{l} 
    \binom{l}{j} 
    \Biggl\{
\N\\ &&
    \sum_{k=1}^{j} 
    \binom{j}{k}
    \sum_{r=0}^{l-k}
    \binom{l-k}{r} (-1)^{l+j+k+r}\; 
    \Gam{k+r+m+n+\eph}{m+1, n+1, k+r+\eph}
\N\\ && 
    \times  
     B\left(k, m+1+ \eph\right) 
    \frac{B\left(k+m-\eph, r+1+n - \eph\right) 
      B\left(r+l-1, n+1 + \eph\right)}
      {(k+r+1+m+n-\ep)(N+3-j)} 
\N\\& & 
  + \sum_{r=0}^{l-j}
    \binom{l-j}{r} (-1)^{l+j+r}\;
    \Gam{j+r+m+n+ \eph}{m+1, n+1, j+r+\eph}
    B\left(j, m+1 + \eph \right) 
\N\\ & & 
    \times
    \frac{ B\left(j+m-\eph,r+1+n-\eph\right)   
     B\left(r+l-1,n+1+\eph\right)}
    {(j+r+1+m+n-\ep) (N+3-j)} 
    \Biggr\}~.   
\end{eqnarray}
\begin{table}[htbp]
  \begin{center}
  \renewcommand{\arraystretch}{1.3} 
  \begin{tabular}{|l|c|l||l|c|l|}
  \hline
  \multicolumn{1}{|c}{Diagram} &
  \multicolumn{1}{|c}{$N$} &
  \multicolumn{1}{|c||}{  } &
  \multicolumn{1}{|c}{Diagram} &
  \multicolumn{1}{|c}{$N$} &
  \multicolumn{1}{|c|}{  } \\
  \hline
  $\hat{I}_{1a}$ & 0 & $2 - 2 \zeta_3$ &
  $\hat{I}_{2b}$  & 0 & $\frac{1}{8} $                     \\
           & 1 & $ 1 - \zeta_3$   &
           & 1 & $ \frac{5}{108}        $                \\
           & 2 & $ \frac{199}{324} -\frac{11}{18} \zeta_3 $ &
           & 2 & $ \frac{731}{1728} - \frac{1}{3} \zeta_3        $ \\
           & 3 & $ \frac{91}{216}  - \frac{5}{12} \zeta_3 $  &
           & 3 & $ \frac{2142253}{5184000} -\frac{1}{3} \zeta_3         $ \\
  \hline
  $\hat{I}_{1b}$ & 0 & $-\frac{9}{4} + 2 \zeta_3  $   &                                     
  $\hat{I}_3$    & 0 & $ 2-2 \zeta_3  $ \\
           & 1 & $-\frac{247}{216} + \zeta_3   $  &                                      
           & 1 & $ 0$            \\
           & 2 & $-\frac{1831}{2592}+\frac{11}{18} \zeta_3$  &                                      
           & 2 & $ \frac{967}{432} - 2\zeta_3$ \\
           & 3 & $-\frac{1257637}{2592000}+\frac{5}{12} \zeta_3   $  &                                      
           & 3 & $ 0$ \\
  \hline 
  $\hat{I}_{2a}$ & 0 & $ 2 - 2 \zeta_3$ &
  $\hat{I}_{4}$ & 0 & $ 2 \zeta_3 $ \\
           & 1 & $ 1 -  \zeta_3  $   &              
           & 1 & $ 2 - 2 \zeta_3   $ \\
           & 2 & $ \frac{1399}{1296} - \zeta_3$  &   
           & 2 & $ \frac{29}{12}   - \frac{83}{36} \zeta_3 $ \\
           & 3 & $ \frac{967}{864}  - \zeta_3 $  &  
           & 3 & $ \frac{17}{6} - \frac{47}{18} \zeta_3$ \\
  \hline
 \end{tabular}
 \caption{Mellin Moments for the Integrals $\hat{I}_{1a}-\hat{I}_{4}$.}
 \label{TAB1}
 \end{center}
\end{table}

These sums are performed by {\sf Sigma} and yield
\begin{eqnarray}
 \hat{I}_{2a} &=& \frac{1}{(N+1) (N+2) (N+3)}\Biggl\{
   2^{N+4} S_{1,2}\left(\frac{1}{2},1\right)
  +2^{N+3} S_{1,1,1}\left(\frac{1}{2},1,1\right)
  +\frac{\left(N^2+12 N+16\right)}{2 (N+1) (N+2)} S_{1}^2
\N\\& &
  +\frac{\left(3 N^2+40 N+56\right)}{2 (N+1) (N+2)} S_{2}
  +\frac{1}{6} S_{1}^3
  +\frac{4 (2 N+3)}{(N+1)^2 (N+2)} S_{1}
  -\frac{1}{2} S_{2} S_{1}-(-1)^N S_{-3}
\N\\& &
  +\frac{1}{3} (-3 N-17) S_{3}
  -2 (-1)^N S_{-2,1}
  +(-N-3) S_{2,1}
  -2 (-1)^N \zeta_3
  -2 \left(2^{N+3}
  -3\right) \zeta_3
\N\\& &
  +\frac{8 (2 N+3)}{(N+1)^3 (N+2)}
   \Biggr\} + O(\ep)~.
\end{eqnarray}
Here generalized harmonic sums emerge with $\xi \in \{\frac{1}{2},1\}$ together with
powers $2^N$. In the limit $N \rightarrow \infty$ the generalized harmonic sums approach
finite values given below. Still $\hat{I}_{2a}$ does not diverge exponentially due to
relations among these special generalized harmonic sums, \cite{ABS1}. The asymptotic
series of $\hat{I}_{2a}$ was computed using {\sf HarmonicSums} \cite{HARMSU} and is 
given by
{\small
\begin{eqnarray}
\hat{I}_{2a}(N) &\simeq& 
\Biggl[\Bigl(\frac{4665}{N^{10}}-\frac{3025}{2 N^9}+\frac{483}{N^8}-\frac{301}{2
N^7}+\frac{45}{N^6}-\frac{25}{2 N^5}+\frac{3}{N^4}-\frac{1}{2 N^3}\Bigr) L(N)
\N\\ &&
-\frac{84311831}{168 N^{10}}+\frac{73442819}{1008 N^9}-\frac{495179}{40
N^8}+\frac{185923}{80 N^7}-\frac{903}{2 N^6}+\frac{2029}{24 N^5}-\frac{55}{4
N^4}+\frac{3}{2 N^3}\Biggr] \zeta_2
\N\\ &&
+\Bigl(-\frac{1555}{N^{10}}+\frac{3025}{6 N^9}-\frac{161}{N^8}+\frac{301}{6
N^7}-\frac{15}{N^6}+\frac{25}{6N^5}-\frac{1}{N^4}+\frac{1}{6 N^3}\Bigr) L^3(N)
\N\\ &&
+
\Bigl(-\frac{15557449}{168 N^{10}}+\frac{12331933}{1008 N^9}-\frac{72181}{40
N^8}+\frac{22877}{80 N^7}-\frac{97}{2 N^6}+\frac{227}{24 N^5}-\frac{9}{4 
N^4}+\frac{1}{2
N^3}\Bigr) L^2(N)
\N\\ &&
+\Bigl(\frac{116332471}{240 N^{10}}-\frac{3304037}{56 N^9}+\frac{470549}{60
N^8}-\frac{1509931}{1440 N^7}+\frac{2897}{24 N^6}-\frac{145}{24 N^5}-\frac{5}{2
N^4}+\frac{1}{N^3}\Bigr) L(N)
\N\\ &&
-\frac{334237263613}{423360 N^{10}}+\frac{1140112957301}{12700800
N^9}-\frac{158319577}{14400 N^8}+\frac{1226359}{900 N^7}-\frac{7771}{48
N^6}+\frac{1429}{72 N^5}
\N\\ &&
-\frac{15}{4N^4}+\frac{1}{N^3}
+\Bigl(-\frac{32633}{N^{10}}+\frac{32545}{3 N^9}-\frac{3601}{N^8}+\frac{3577}{3
N^7}-\frac{393}{N^6}+\frac{385}{3N^5}-\frac{41}{N^4}
\N\\ &&
+\frac{37}{3 
N^3}-\frac{3}{N^2}\Bigr)
\zeta_3 +O\left(\frac{1}{N^{11}}\right)~,
\end{eqnarray}
}
where $L(N) = \ln(N) + \gamma_E$. Likewise, one obtains for $I_{2b}$
\begin{eqnarray}
 I_{2b} &\equiv& \frac{i(\Delta.p)^Na_s^3S_\ep^3}{(m^2)^{3-3\eph}}\hat{I}_{2b},
\\
 \hat{I}_{2b} 
 &=&   
    -\exp\left(-3\eph\gamma_E\right)
     \Gamma\left(3 - \frac{3}{2}\ep\right) 
     \frac{1}{(N+1) (N+2)}
     \sum_{m=0}^{\infty}
     \sum_{n=0}^{\infty}
     \sum_{l=2}^{N+2}
     \binom{N+2}{l}
     \sum_{j=2}^{l}
     \binom{l}{j} 
     \Biggl[
\N
\end{eqnarray}
\begin{eqnarray}
&&
     \phantom{\times}
     \sum_{k=1}^{j}
     \binom{j}{k}
     \sum_{r=0}^{l-k}
     \binom{l-k}{r} 
     (-1)^{l+j+k+r}\;
     \Gam{k+r+m+n+ \eph}{m+1, n+1, k+r+\eph}
     B\left(k, m+1+ \eph\right)
\N
\\
&& 
\times  
     B(N+3-j,2) 
     \frac{ B\left(k+m-\eph, r+2+n - \eph\right) 
       B\left(r+l-1, n+1 + \eph\right)}{k+r+2+m+n-\ep} 
\N\\ & & 
    +\sum_{r=0}^{l-j}
     \binom{l-j}{r} 
     (-1)^{l+j+r}\;
     \Gam{j+r+m+n+ \eph}{m+1, n+1, j+r+\eph}
     B\left(j, m+1 + \eph \right)
\N
\end{eqnarray}
\begin{eqnarray}
& & 
     \times
     B(N+3-j,2)
     \frac{B\left(j+m-\eph,r+2+n-\eph\right)   
       B\left(r+l-1,n+1+\eph\right)}{j+r+2+m+n-\ep} \Biggr\}   
\\
&=& \frac{1}{(N+1) (N+2) (N+3) (N+4)}
   \Biggl\{
    2^{N+4} N S_{1,2}\left(\frac{1}{2},1\right)
   +2^{N+3} N S_{1,1,1}\left(\frac{1}{2},1,1\right)
\N\\& &
   +(-1)^N (N^2+4 N+2)\left(
     -S_{-3}
     -2 S_{-2,1}
     -2 \zeta_3
   \right)
   +\frac{1}{3} \left(-6 N^2-33 N-20\right) S_{3}
\N\\& &
   +\frac{\left(8 N^3+31 N^2+17 N-18\right)}{2 (N+1) (N+2) (N+3)} S_{1}^2
   +\frac{2 \left(9 N^3+43 N^2+58 N+21\right)}{(N+1)^2 (N+2) (N+3)} S_{1}
\N\\& &
   +\frac{3 \left(12 N^3+55 N^2+61 N+6\right)}{2 (N+1) (N+2) (N+3)} S_{2}
   -\frac{1}{3} S_{1}^3
   +S_{2} S_{1}
   +(N+4) S_{2,1}
\N\\& &
   +2 \left(N^2-2^{N+3} N+6 N+2\right) \zeta_3
\N\\& &
   +\frac{-6 N^5-14 N^4+68 N^3+247 N^2+225 N+36}{(N+1)^3 (N+2)^2 (N+3)}
   \Biggr\}  + O(\ep)~.
\end{eqnarray}
Again the sum-structure of the integrals remains the same.
Unlike the case for the massless 3-loop Wilson coefficients~\cite{MVV2} and massive 
integrals in \cite{Ablinger:2010ty} the generalized harmonic sums do not vanish 
diagram by diagram. We remark that 
sums of this type even emerge in massive 2-loop integrals, if diagrams are simply separated
into individual terms in a mathematical manner, e.g. in a fully automated computation 
to $O(\varepsilon)$ \cite{BBK4}, while they are absent in case the diagrams are considered 
as whole entities being mapped to various final sums \cite{Bierenbaum:2007qe,BBK4}. 
The presence of these generalized harmonic sums does not alter the structure of the 
diagrams significantly in the special way they appear, as we will outline below.

For diagram~3 a simple representation is obtained~:
\begin{eqnarray}
I_{3} &=&   \left[1 + (-1)^N\right] I_{2a}(N+1)~.
\end{eqnarray}
Diagrams with a gluon-quark-quark operator insertion on an external gluon 
line can always be related to a diagram with the operator insertion on the fermion 
lines next to this vertex due to the Feynman rule for the operators \cite{BBK2}, 
cf. also Appendix \ref{app:FR} and \cite{BBK5}.

Let us now turn to diagram $I_4$, 
\begin{eqnarray}
I_{4} &\equiv& \frac{i(\Delta.p)^Na_s^3S_\ep^3}{(m^2)^{2-3\eph}}\hat{I}_{4}~.
\end{eqnarray}
We perform the calculation of this diagram in a different way than used before and 
apply the $\alpha$-representation for the Feynman parameters, cf. \cite{ALPHAP}.
The integral is convergent in the limit $\ep \rightarrow 0$ and shall be calculated 
using the method of hyperlogarithms. This method has been worked out in \cite{BROWN}
in case the numerator functions are fixed polynomials, the integrands are
rational functions in the parameters $\alpha_i$ for convergent integrals in $D=4$ 
dimensions. It can be applied if an integration order can be found, such that at each
integration step the denominator factors into linear polynomials in the respective
integration variable.
 
Integral $\hat{I}_4$ is given by~:
\begin{eqnarray}
\hat{I}_4 &=& \int \cdots \int_{\alpha_i>0,\alpha_1+\alpha_2=1} \prod_{i=1}^7 d\alpha_i 
\frac{\sum_{j=0}^{N}{T_{4a}^{N-j} T_{4b}^j}}{U_G^{2} M_G^{2}}, 
\end{eqnarray}
with the polynomials, cf.~ \cite{ALPHAP},
\begin{eqnarray}
T_{4a}&=&
\alpha_{5} \alpha_{7} \alpha_{4}+\alpha_{2} \alpha_{3}
\alpha_{5}+\alpha_{2} \alpha_{5} \alpha_{4}+\alpha_{3} \alpha_{5}
\alpha_{7}+\alpha_{2} \alpha_{5} \alpha_{8}+\alpha_{8} \alpha_{5}
\alpha_{4}+\alpha_{5} \alpha_{7} \alpha_{8}+\alpha_{2} \alpha_{3}
\alpha_{8}
\N\\&&
+\alpha_{7} \alpha_{2} \alpha_{8}+\alpha_{6} \alpha_{2} \alpha_{8}+\alpha_{3} \alpha_{7} \alpha_{2}+\alpha_{2} \alpha_{3} \alpha_{6}+\alpha_{4} \alpha_{2} \alpha_{8}+\alpha_{2} \alpha_{6} \alpha_{4}+\alpha_{4} \alpha_{7} \alpha_{2}
\end{eqnarray}
\begin{eqnarray}
T_{4b}&=&\alpha_{2} \alpha_{5} \alpha_{4}+\alpha_{4} \alpha_{2} \alpha_{8}+\alpha_{4} \alpha_{7} \alpha_{2}+\alpha_{2}
\alpha_{5} \alpha_{8}+\alpha_{2} \alpha_{3} \alpha_{5}+\alpha_{7} \alpha_{2} \alpha_{8}+\alpha_{3} \alpha_{7}
\alpha_{2}
+\alpha_{8} \alpha_{5} \alpha_{4}
\N\\&&
+\alpha_{5} \alpha_{7} \alpha_{4}
+\alpha_{4} \alpha_{1} \alpha_{8}+\alpha_{1} \alpha_{7} \alpha_{4}+\alpha_{3} \alpha_{5} \alpha_{7}
+\alpha_{5} \alpha_{7} \alpha_{8}+\alpha_{8} \alpha_{1} \alpha_{7}+\alpha_{1} \alpha_{3} \alpha_{7}\\
U_G&=&
\alpha_{2} \alpha_{5} \alpha_{4}+\alpha_{2} \alpha_{3} \alpha_{5}+\alpha_{1} \alpha_{3} \alpha_{5}
+\alpha_{5} \alpha_{7}
\alpha_{4}+\alpha_{1} \alpha_{6} \alpha_{4}+\alpha_{1} \alpha_{3} \alpha_{6}+\alpha_{2} \alpha_{3} \alpha_{6}+\alpha_{2}
\alpha_{6} \alpha_{4}
\N\\&&
+\alpha_{5} \alpha_{6} \alpha_{4}
+\alpha_{1} \alpha_{5} \alpha_{4}+\alpha_{3} \alpha
_{5} \alpha_{7}+\alpha_{1} \alpha_{3} \alpha_{7}+\alpha_{1} \alpha_{7}
\alpha_{4}
+\alpha_{3} \alpha_{7} \alpha_{2}+\alpha_{4} \alpha_{7} \alpha_{2}+\alpha_{3} \alpha_{5}
 \alpha_{6}
\N\\&&
+\alpha_{2} \alpha_{3} \alpha_{8}
+\alpha_{2} \alpha_{5} \alpha_{8}
+\alpha_{5} \alpha_{7} \alpha_{8}+\alpha_{8} \alpha_{5} \alpha_{4}+\alpha_{8} \alpha_{5} \alpha_{6}+\alpha_{5} \alpha_{3}
\alpha_{8}+\alpha_{1} \alpha_{8} \alpha_{5}+\alpha_{1} \alpha_{8}
\alpha_{6}
\N\\&&
+\alpha_{6} \alpha_{2} \alpha_{8}
+\alpha_{1}
\alpha_{8} \alpha_{3}
+\alpha_{4} \alpha_{1} \alpha_{8}+\alpha_{4} \alpha_{2} \alpha_{8}+\alpha_{7} \alpha_{2} \alpha_{8}
+\alpha_{8} \alpha_{1} \alpha_{7}\\
M_G&=& \alpha_{1} + \alpha_{2}
+ \alpha_{3}
+ \alpha_{4}
+ \alpha_{6}
+ \alpha_{7}~.
\end{eqnarray}
In order to tackle diagram 4 with the method of hyperlogarithms in case of general values of $N$, the 
following ideas have been incorporated. We first perform the transformation
\begin{eqnarray}
\sum_{j=0}^{N} T_{4a}^{N-j} T_{4b}^j &\rightarrow& \sum_{N=0}^\infty \sum_{j=0}^{N} x^N T_{4a}^{N-j} T_{4b}^j
= \sum_{N=0}^\infty \frac{
  (T_{4a} x)^N
- (T_{4b} x)^N}{T_{4a} - T_{4b}} \N\\ &=& \frac{1}{T_{4a} - T_{4b}} \left[
  \frac{1}{1 - xT_{4a}} 
- \frac{1}{1 - xT_{4b}} \right] 
= \frac{x}{(1- x T_{4a})(1- x T_{4b})}~, 
\label{eq:XOP}
\end{eqnarray}
constructing a formal power series being resummed in the tracing parameter $x$.  This 
effectively moves the action of the local operators into propagator--like terms. For more 
complicated operator insertions the product structure of the denominator in (\ref{eq:XOP}) 
is simply extended. A transformation of this type assumes that one finally can map the 
function $\tilde{f}(x)$ being obtained back to the desired solution $f(N)$ directly. 

We consider the following iterated integrals $\LL_{\vec{a}}(x) \equiv \LL_{\vec{a}}$
\begin{eqnarray}
\LL_{b,\vec{a}}(x) = \int_0^x \frac{dy}{y-b} \LL_{\vec{a}}(y),~~~\LL_{\emptyset}(y) = 
1;~~~
\LL_{\underbrace{\mbox{\scriptsize 0, \ldots ,0}}_{\mbox{\scriptsize
$n$}}}(x)  = \frac{1}{n!} \ln^n(x)~.
\end{eqnarray}
During the integration process the indices $a_i$ are usually rational functions of $x$ 
and the integration variables $\alpha_i$. Because of this the iterated integrals
$\LL_{\vec{a}}$ are called hyperlogarithms rather than polylogarithms over the alphabet
$\{a_1, \ldots, a_k\}$.

\begin{eqnarray} 
\tilde{I}_4(x) &=& \Biggl[ - \frac{1+x}{x^3} \LL_{-1} - \frac{2x-1}{x^3} \LL_{1/2} - \frac{3(1-x)}{x^3} 
\LL_{1} - \frac{1-2x+x^2}{(1-x)x^3} \LL_{0,-1} + \frac{1 - 2 x^2}{x^3} \LL_{0,1/2} 
\N\\ && 
-\frac{3-4x-3x^2+3x^3}{(1-x)x^3} \LL_{0,1} -\frac{1-2x^2}{x^3} \LL_{1,1/2} +\frac{(1-x)(2+3x)}{x^3} 
\LL_{1,1} \Biggr] \zeta_3 
\N\\ 
&& +\frac{(1+x) }{2 x^3} \left(3 
\LL_{-1,0,0,1} - 2 \LL_{-1,0,1,1} - 3 \LL_{1,0,0,1}\right) +\frac{1}{x^2} \left(6 \LL_{0,0,1,1} - 4 
\LL_{0,1,0,1} - \LL_{0,1,1,1} \right) 
\N
\end{eqnarray} 
\begin{eqnarray}  
&& -\frac{(-1+2 x) }{2 x^3} \left[ 3 \LL_{1/2,0,0,1} - 
\LL_{1/2,0,1,1} - 3 \LL_{1/2,1,0,1} + \LL_{1/2,1,1,1} \right] \N\\ && -\frac{3 }{2 x^2} \LL_{1,0,1,1} 
+\frac{2 }{x^2} \LL_{1,1,0,1} -\frac{(-1+x) }{2 x^3} \LL_{1,1,1,1} +\frac{2 }{x^2} \left(\LL_{0,1,1} - 
\LL_{1,0,1} \right) \N\\ && +\frac{\left(-1+2 x+x^2\right) }{2 (-1+x) x^3} \left[ 3 \LL_{0,-1,0,0,1} - 2 
\LL_{0,-1,0,1,1} \right] \N\\ && -\frac{5 }{-1+x} \LL_{0,0,0,1,1} -\frac{5 }{2 (-1+x)} \LL_{0,0,1,0,1} 
+\frac{3 (3+x) }{2 (-1+x) x} \LL_{0,0,1,1,1} \N\\ && -\frac{\left(-1+2 x^2\right) }{2 x^3} \left[3 
\LL_{0,1/2,0,0,1} + \LL_{0,1/2,0,1,1} + 3 \LL_{0,1/2,1,0,1} - \LL_{0,1/2,1,1,1} \right] 
\N\\ 
&& +\frac{3 \left(1-3 x^2+3 x^3\right) }{2 (-1+x) x^3} \LL_{0,1,0,0,1} +\frac{8-14 
x+5 x^2+3 x^3}{2 (-1+x) x^3} \LL_{0,1,0,1,1} \N\\ && +\frac{8-15 x+3 x^2}{2 (-1+x) x^3} \LL_{0,1,1,0,1} 
-\frac{3 (-3+2 x) }{2 x^3} \LL_{0,1,1,1,1} +\frac{-6+3 x+5 x^2}{x^3} \LL_{1,0,0,1,1} \N\\ && +\frac{2 
(-1+x)}{x^3} \LL_{1,1,1,0,1} +\frac{4-2 x+5 x^2}{2 x^3} \LL_{1,0,1,0,1} -\frac{-4+6 x+3 x^2}{2 x^3} 
\LL_{1,0,1,1,1} \N\\ && +\frac{\left(-1+2 x^2\right) }{2 x^3} \left[ 3 \LL_{1,1/2,0,0,1} - 
\LL_{1,1/2,0,1,1} \right] -\frac{3 (-1+x) (4+3 x)}{2 x^3} \LL_{1,1,0,0,1} \N\\ && -\frac{\left(-1+2 
x^2\right)}{2 x^3}
  \left[ \LL_{1,1/2,1,0,1}
+  \LL_{1,\frac{1}{2},1,1,1} \right]
-\frac{(-1+x) (5+3 x) }{2 x^3} \LL_{1,1,0,1,1}~.
\label{eq:LL4}
\end{eqnarray}

One first calculates the result of this transformation for integral $\hat{I}_{4}$ in terms of the variable $x$.
It can be expressed in hyperlogarithms $\LL_{\vec{a}}(x)$ over the alphabet 
$\{0,1,-1,1/2\}$ \footnote{For more complicated diagrams also the parameter $x$ appears in terms 
of rational functions in the alphabet.}  which are thus
generalized harmonic polylogarithms \cite{MUW,ABS1}~: 
Finally the transformation (\ref{eq:XOP}) is reversed finding the $N$th 
expansion
coefficient of (\ref{eq:LL4}) symbolically using the packages {\sf Sigma} 
\cite{SIGMA} and {\sf HarmonicSums} \cite{HARMSU}. One obtains in $N$-space 
the following representation in terms of harmonic sums and their generalizations~:
\begin{eqnarray}
\hat{I}_4 &=&
\frac{P_1}{2 (1+N)^5 (2+N)^5 (3+N)^5}
+\frac{P_2}{(1+N)^2 (2+N)^2 (3+N)^2} \zeta_3
\N\\&&
+\frac{(-1)^N \left(65+101 N+56 N^2+13 N^3+N^4\right) }{2
  (1+N)^2 (2+N)^2 (3+N)^2} S_{-3}
+\frac{\left(-24-5 N+2 N^2\right) }{12 (2+N)^2 (3+N)^2} S_1^3
\N\\&&
-\frac{1}{2 (1+N) (2+N) (3+N)} S_2^2
+\frac{1}{(2+N) (3+N)} S_1^2 S_2
\N\\&&
+ \frac{314+631 N+578 N^2+288 N^3+68 N^4+5 N^5}{4 (1+N)^3 (2+N)^2
    (3+N)^2} S_1^2
-\frac{3}{2} S_5
\N\\&&
-\frac{\left(399+2069 N+2774 N^2+1510 N^3+349 N^4+27 N^5\right)}{6 (1+N)^2
  (2+N)^2 (3+N)^2} S_3
-2 S_{-2,-3}
\N\\&&
-2 \zeta_3 S_{-2} 
-S_{-2,1} S_{-2}
+\frac{(-1)^N \left(65+101 N+56 N^2+13
    N^3+N^4\right)}{(1+N)^2 (2+N)^2 (3+N)^2} S_{-2,1}
\N\\&&
+\frac{\left(59+42 N+6 N^2\right)}{2 (1+N)  (2+N) (3+N)} S_4
+\frac{(5+N)}{(1+N) (3+N)} \zeta_3 S_{1}\Bigl(2\Bigr)
\N\\&&
-\frac{752+2087 N+2490 N^2+1580 N^3+558 N^4+105 N^5+8 N^6}{4 (1+N)^3 (2+N)^2
    (3+N)^2} S_2
-\zeta_3 S_2
\N
\end{eqnarray}
\begin{eqnarray}
&&
-\frac{3}{2} S_3 S_2
-2 S_{2,1} S_2
+\frac{\left(99+225 N+190 N^2+65 N^3+7 N^4\right)}{2 (1+N)^2
  (2+N)^2 (3+N)} S_{2,1}
\N\\&&
+\frac{P_3}{(1+N)^4 (2+N)^4
    (3+N)^4} S_1 
-\frac{(11+5 N) }{(1+N) (2+N) (3+N)} \zeta_3 S_1 
\N\\&&
-\frac{\left(470+1075 N+996 N^2+447 N^3+96 N^4+8 N^5\right)}{4 (1+N)^2
  (2+N)^2 (3+N)^2}  S_2 S_1 
-S_{2,3}
\N\\
&&
+\frac{(53+29 N)}{2 (1+N) (2+N) (3+N)}  S_3 S_1 
-\frac{3 (3+2 N)}{(1+N) (2+N) (3+N)} S_1 S_{2,1}
\N\\&&
+\frac{\left(-79-40 N+N^2\right)}{2 (1+N) (2+N) (3+N)} S_{3,1}
-3 S_{4,1}
+S_{-2,1,-2}
\N\\&&
+\frac{2^{1+N} \left(-28-25 N-4 N^2+N^3\right)}{(1+N)^2 (2+N)
  (3+N)^2} S_{1,2}\left(\frac{1}{2},1\right)
-\frac{\left(-7+2 N^2\right)}{(1+N) (2+N) (3+N)} S_{2,1,1}
\N\\
&&
+5 S_{2,2,1}
+6 S_{3,1,1}
+\frac{2^N \left(-28-25 N-4 N^2+N^3\right)}{(1+N)^2 (2+N)
  (3+N)^2}  S_{1,1,1}\left(\frac{1}{2},1,1\right)
\N\\
&&
-\frac{(5+N)}{(1+N) (3+N)} S_{1,1,2}\left(2,\frac{1}{2},1\right)
-\frac{(5+N) }{2 (1+N) (3+N)} S_{1,1,1,1}\left(2,\frac{1}{2},1,1\right)~,
\end{eqnarray}
where
\begin{eqnarray}
P_1&=&-31104-159408 N-353808 N^2-446652 N^3-353808 N^4-182604
  N^5-61488 N^6
\N\\&& -13044 N^7-1584 N^8-84 N^9,
\\
P_2&=&-105+65 (-1)^N+7\ 2^{4+N}-150 N+101 (-1)^N N+39\ 2^{2+N} N-73 N^2+56 (-1)^N 
N^2
\N\\&& +33\ 2^{1+N} N^2-12 N^3+13 (-1)^N N^3+2^{2+N} N^3+(-1)^N N^4-2^{1+N} N^4,
\\
P_3&=&5436+29004 N+67285 N^2+89175 N^3+74616 N^4+41120 N^5+15107 N^6+3659 N^7
\N\\&& +562 N^8+50 N^9+2 N^{10}~.
\end{eqnarray}

In order to show that the $2^N$-terms are artifacts of the
representation as in $I_{2a,b}$, we calculate the asymptotic representation of 
$\hat{I}_4$~: 
{\small
\begin{eqnarray}
 \hat{I}_4 &\simeq&
  \left[
   \frac{1115231}{20 N^{10}}
  -\frac{74121}{4 N^9}
  +\frac{122951}{20 N^8}
  -\frac{40677}{20 N^7}
  +\frac{13391}{20 N^6}
  -\frac{873}{4 N^5}
  +\frac{1391}{20 N^4}
  -\frac{417}{20 N^3}
  +\frac{101}{20 N^2}
  \right] \zeta_2^2
\N\\&&
+
  \Biggl[
   \left(
   -\frac{95855}{2 N^{10}}
   +\frac{31525}{2 N^9}
   -\frac{10295}{2 N^8}
   +\frac{3325}{2 N^7}
   -\frac{1055}{2 N^6}
   +\frac{325}{2 N^5}
   -\frac{95}{2 N^4}
   +\frac{25}{2 N^3}
   -\frac{5}{2 N^2}
   \right) L(N)
\N\\&&
  -\frac{23280115}{2016 N^{10}}
  +\frac{2093041}{1008 N^9}
  -\frac{177251}{1008 N^8}
  -\frac{25843}{336 N^7}
  +\frac{2569}{48 N^6}
  -\frac{155}{8 N^5}
  +\frac{91}{24 N^4}
  +\frac{2}{3 N^3}
  -\frac{11}{12 N^2}
  \Biggr] \zeta_3
\N\\&&
+
 \Biggl[
  \left(
   \frac{19171}{N^{10}}
  -\frac{6305}{N^9}
  +\frac{2059}{N^8}
  -\frac{665}{N^7}
  +\frac{211}{N^6}
  -\frac{65}{N^5}
  +\frac{19}{N^4}
  -\frac{5}{N^3}
  +\frac{1}{N^2}
  \right)L^2(N)
\N\\&&
 +\left(
   \frac{103016863}{2520 N^{10}}
  -\frac{3091261}{315 N^9}
  +\frac{2571839}{1260 N^8}
  -\frac{6215}{21 N^7}
  -\frac{293}{20 N^6}
  +\frac{2071}{60 N^5}
  -\frac{103}{6 N^4}
  +\frac{67}{12 N^3}
  -\frac{1}{N^2}
  \right) L(N)
\N\\&&
 +\frac{292993001621}{302400 N^{10}}
 -\frac{4402272031}{30240 N^9}
 +\frac{22261739}{840 N^8}
 -\frac{78507473}{14112 N^7}
 +\frac{180961}{144 N^6}
 -\frac{111807}{400 N^5}
 +\frac{629}{12 N^4}
\N\\&&
 -\frac{319}{72 N^3}
 -\frac{7}{4 N^2}
 \Biggr] \zeta_2
\N
\end{eqnarray}
\begin{eqnarray}
&&
 +\left(
   \frac{249223}{6 N^{10}}
  -\frac{145015}{12 N^9}
  +\frac{10295}{3 N^8}
  -\frac{11305}{12 N^7}
  +\frac{1477}{6 N^6}
  -\frac{715}{12 N^5}
  +\frac{38}{3 N^4}
  -\frac{25}{12 N^3}
  +\frac{1}{6 N^2}
  \right) L^3(N)
\N\\&&
 +\Bigl(
   \frac{193493767}{10080 N^{10}}
  +\frac{210658237}{10080 N^9}
  -\frac{21541697}{2520 N^8}
  +\frac{243269}{96 N^7}
  -\frac{30539}{48 N^6}
  +\frac{2123}{16 N^5}
  -\frac{59}{3 N^4}
\N\\&&
  +\frac{5}{8 N^3}
  +\frac{1}{2 N^2}
  \Bigl) L^2(N)
\N\\&&
 +\Bigl(
  -\frac{2207364771673}{4233600 N^{10}}
  +\frac{1390655509}{352800 N^9}
  +\frac{285594061}{22050 N^8}
  -\frac{67234111}{14400 N^7}
  +\frac{8617073}{7200 N^6}
  -\frac{35209}{144 N^5}
\N\\&&
  +\frac{116}{3 N^4}
  -\frac{119}{24 N^3}
  +\frac{1}{N^2}
  \Bigr) L(N)
\N\\&&
 +\frac{1344226725047831}{889056000 N^{10}}
 -\frac{165849841805771}{889056000 N^9}
 +\frac{808151260279}{27783000 N^8}
 -\frac{708430537}{120960 N^7}
 +\frac{304474703}{216000 N^6}
\N\\&&
 -\frac{606811}{1728 N^5}
 +\frac{1867}{24 N^4}
 -\frac{1813}{144 N^3}
 +\frac{1}{N^2} + O\left(\frac{1}{N^{11}}\right)~,
\end{eqnarray}
}
which shows a convergent behaviour.

In the above expressions generalized harmonic sums occur which are convergent
in the limit $N \rightarrow \infty$. Furthermore, all of them can be represented
in terms of multiple zeta values~:
\begin{eqnarray}
S_1\left(\frac{1}{2};\infty \right) 
&=&\ln (2)
\\
S_2\left(\frac{1}{2};\infty \right)
&=&\frac{1}{2} \left[\zeta_2-\ln ^2(2)\right]
\\
S_{1,1}\left(\frac{1}{2},1;\infty \right)
&=&\frac{1}{2} \zeta_2
\\
S_3\left(\frac{1}{2};\infty \right)
&=&\frac{1}{6} \ln ^3(2) - \frac{1}{2} \zeta_2 \ln(2) + \frac{7}{8} \zeta_3
\\
S_{1,2}\left(\frac{1}{2},1;\infty \right)
&=&\frac{5}{8} \zeta_3
\\
S_{2,1}\left(1,\frac{1}{2};\infty \right)
&=&\frac{1}{2}\zeta_2 \ln (2) +\frac{1}{4} \zeta_3 + \frac{1}{6}  \ln^3(2)
\\
S_{2,1}\left(\frac{1}{2},1;\infty \right)
&=&\zeta_3-\frac{1}{2} \zeta_2 \ln(2)
\\
S_{2,1}\left(\frac{1}{2},2;\infty \right)
&=&\frac{21}{8} \zeta_3 - \frac{3}{2} \zeta_2 \ln(2)
\\
S_{1,2}\left(\frac{1}{2},2;\infty \right)
&=&\frac{3}{2} \zeta_2 \ln (2)
\\
S_{1,1,1}\left(\frac{1}{2},1,1;\infty \right)
&=&\frac{3 \zeta_3}{4}
\\
S_{1, 1, 1}\left(\frac{1}{2}, 2, 1; \infty \right)  
&=& \frac{3}{2} \ln(2) \zeta_2 + \frac{7}{4} \zeta_3~.
\end{eqnarray}
For completeness, we also give the representations of the generalized harmonic sums
in terms of a Mellin transformation
\begin{eqnarray}
\Mvec[f(x)](N) = \int_0^1 dx x^{N-1}~f(x),~~\text{resp.}~~
\Mvec\left[\left(\frac{g(x)}{1-x}\right)_+\right](N) = \int_0^1 dx
\frac{x^{N-1}-1}{1-x}~g(x)~.
\end{eqnarray}
It turns out that all cases are related to the usual harmonic polylogarithms
over the alphabet $\{-1,0,1\}$ \cite{VR}, 
adding in a few cases the letters $1/(\frac{1}{2} - x)$ or
$1/(2-x)$ as the first one, for $x \in [0,1]$. Here the second letter is uncritical.
In case of the first letter a new type of $+$-prescription for the Mellin-transform 
emerges on $[0,1]$, through 
which the singularity of the denominator at $x = 1/2$ is regulated, 
\begin{eqnarray}
\Mvec_1\left[\left(\frac{g(x)}{\frac{1}{2}-x}\right)_+\right](N) = \int_0^1 dx 
\frac{(2x)^{N-1}-1}{\frac{1}{2} - x}~g(x)~.
\end{eqnarray}
The corresponding Mellin transforms which represent the generalized harmonic sums are~:
\begin{eqnarray}
\label{eqLMT1}
  S_1(2;N)&=&-\int_0^1 \frac{(2 x)^N-1}{\frac{1}{2}-x} \, dx
\\
    S_{1,2}\left(\frac{1}{2},1;N\right)&=&
    \frac{5}{8} \zeta_3
    + \int_0^1 \frac{\left(\frac{x}{2}\right)^N}{2-x} \HH_{1,0}(x) \, dx
\\
  S_{1,1,1}\left(\frac{1}{2},1,1;N\right)&=&
\frac{3}{4} \zeta_3
    - \int_0^1 \frac{\left(\frac{x}{2}\right)^N}{2-x} \HH_{1,1}(x) \, dx
\end{eqnarray}
\begin{eqnarray}
  S_{1,1,2}\left(2,\frac{1}{2},1;N\right)&=&
     -\frac{5}{8}\zeta_3 \int_0^1 dx \frac{(2x)^N-1}{\frac{1}{2}-x} 
     - \int_0^1\frac{x^N-1}{1-x}H_{-1,0,1}(1-x) dx \N\\ &&
     + \zeta_2 \int_0^1 dx \frac{x^N-1}{1-x}H_{-1}(1-x)  
\\
\label{eqLMT2}
  S_{1,1,1,1}\left(2,\frac{1}{2},1,1;N\right)&=&
-\frac{3}{4} \zeta_3 \left[
\int_0^1 dx \frac{(2x)^N-1}{\frac{1}{2} - x }
- \int_0^1 dx 
\frac{x^N-1}{1-x} 
\right]
\N\\ &&
- \int_0^1 dx \frac{x^N-1}{1-x} \HH_{2,1,1}(x)~,
\end{eqnarray}
with $\HH_{-1}(1-x) = \ln(2 - x) = - \int_0^x dy/(2-y) + \ln(2)$. Similarly, also
$\HH_{-1,0,1}(1-x)$ is a polylogarithm over the alphabet $\{1/x,1/(1-x),1/(2-x)\}$. 

Generalized harmonic sums also occurred in diagram~2. Let us consider the associated
functional $\tilde{I}_{2a}$ in similarity to (\ref{eq:LL4}),
\begin{eqnarray}
\tilde{I}_{2a}(x)&=&
\frac {1}{(N+1) (N+2) (N+3)}
\Biggl\{
-\frac{6 x^3 \zeta_3}{(1-x) (1+x) (\frac{1}{2}- x)}
+\frac{(-2+x) x^2 }{(1-x)^2} \zeta_2 \LL_{1}
\N\\&&
+\frac{2(-2+x) x^2 \left(1-x+x^2\right)}{(1-x)^2 (1+x) (\frac{1}{2}-x)}  \LL_{0,0,1}
-\frac{x \left(-4+4 x+13 x^2-17 x^3+2 x^4\right)}{2 (1-x)^2 (1+x) (\frac{1}{2}-
  x)}  \LL_{0,1,1}
\N\\&&
+\frac{ x \left(5-8 x+2 x^2\right)}{(1-x) (\frac{1}{2}- x)}  \LL_{1,0,1}
-\frac{x \left(7-13 x+2 x^2\right)}{2 (1-x) (\frac{1}{2} - x)} \LL_{1,1,1}
\Biggr\}~.
\label{eq:LL2a}
\end{eqnarray}
Here all harmonic polylogarithms are formed over the alphabet $\{0,1\}$
and the letter $1/(\frac{1}{2}-x)$ only emerges as pre-factor. It causes the generalized 
sums in $I_{2a}(N)$. The structures beyond the usual harmonic polylogarithms in 
(\ref{eq:LL2a}) are much simpler than in (\ref{eq:LL4}), which explains the 
greater difficulty to solve $I_4$ using summation technologies.
\subsection{Diagrams with Three Fermion Propagators}
\label{sec:3.2}

\begin{figure}[htbp]
 \centering
 \parbox[c]{\textwidth}{
  \centering\epsfig{figure=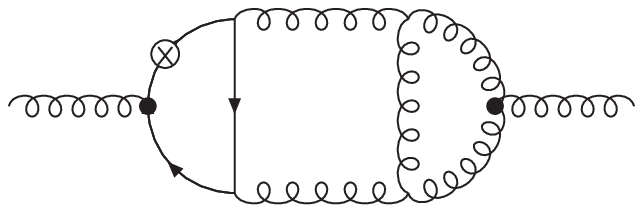,width=0.27\linewidth} \hspace{3mm}
  \centering\epsfig{figure=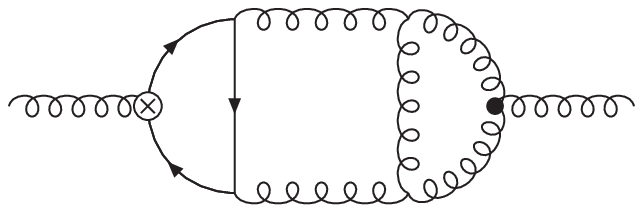,width=0.27\linewidth} \hspace{3mm}
  \centering\epsfig{figure=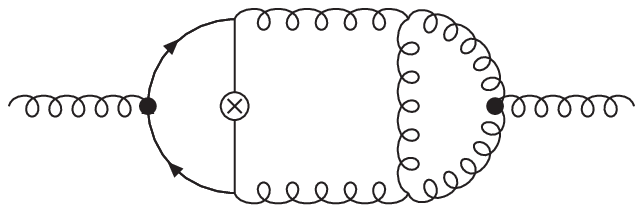,width=0.27\linewidth} \hspace{3mm}
  \\
  {\small 5 \hspace{0.28\textwidth} 6 \hspace{0.28\textwidth} 7}
  \\
  \centering\epsfig{figure=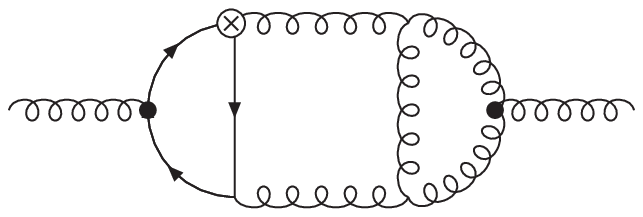, width=0.27\linewidth} \hspace{3mm}
  \centering\epsfig{figure=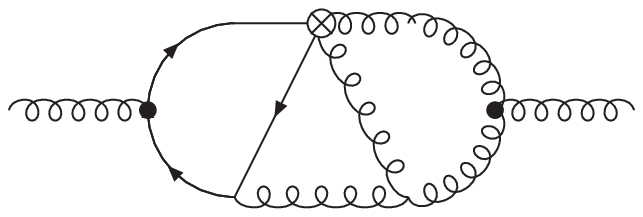,width=0.27\linewidth}
  \\
  {\small 8 \hspace{0.28\textwidth} 9}
 }
 \caption{Diagrams with three fermion propagators}
 \label{fig:Diag3fp}
\end{figure}

\vspace{1mm}
\noindent
The corresponding diagrams are given in Figure~\ref{fig:Diag3fp}.
Again we first give a number of fixed moments in Table~2. Unlike the former 
case, the diagrams contain poles up to $1/\ep^2$.
\begin{table}[htbp]
  \begin{center}
  \renewcommand{\arraystretch}{1.3} 
  \begin{tabular}{|l|c|l|}
  \hline
  \multicolumn{1}{|c}{Diagram} &
  \multicolumn{1}{|c}{N} &
  \multicolumn{1}{|c||}{  } \\
  \hline
  $\hat{I}_{5a}$ & 0 & $  \frac{1}{6} \frac{1}{\ep^2}
                  +\frac{1}{9} \frac{1}{\ep}
                  +\frac{13}{54}+\frac{1}{16} \zeta_2$ \\
           & 1 & $ \frac{1}{12} \frac{1}{\ep^2}
                  +\frac{1}{18} \frac{1}{\ep}
                  +\frac{13}{108}+\frac{1}{32} \zeta_2   $ \\
           & 2 & $ \frac{13}{270} \frac{1}{\ep^2}
                  +\frac{121}{3600} \frac{1}{\ep}
                  +\frac{138911}{1944000}+\frac{13}{720} \zeta_2  $ \\
           & 3 & $ \frac{11}{360} \frac{1}{\ep^2}
                  +\frac{163}{7200} \frac{1}{\ep}
                  +\frac{60911}{1296000}+\frac{11}{960} \zeta_2   $ \\
  \hline
  $\hat{I}_{5b}$ & 0 & $-\frac{1}{10} \frac{1}{\ep^2}
                  +\frac{1}{600} \frac{1}{\ep}
                  -\frac{869}{9000} -\frac{3}{80} \zeta_2 $ \\
           & 1 & $-\frac{1}{24} \frac{1}{\ep^2}
                  +\frac{1}{180} \frac{1}{\ep}
                  -\frac{223}{5400}-\frac{1}{64} \zeta_2 $ \\
           & 2 & $-\frac{13}{630} \frac{1}{\ep^2} 
                  +\frac{127}{33075} \frac{1}{\ep}
                  -\frac{2371837}{111132000}-\frac{13}{1680} \zeta_2 $\\
           & 3 & $-\frac{11}{960} \frac{1}{\ep^2} 
                  +\frac{1919}{806400}\frac{1}{\ep}
                  -\frac{8361911}{677376000} - \frac{11}{2560} \zeta_2 $ \\
  \hline
  $\hat{I}_6$    & 0 & $\frac{1}{6} \frac{1}{\ep^2}
                  +\frac{1}{9} \frac{1}{\ep}
                  +\frac{13}{54} + \frac{1}{16} \zeta_2 $\\
           & 1 & $0$ \\
           & 2 & $\frac{11}{180} \frac{1}{\ep^2}
                  +\frac{163}{3600} \frac{1}{\ep}
                  +\frac{60911}{648000}+\frac{11}{480} \zeta_2 $ \\
           & 3 & $ 0 $ \\
  \hline
  $\hat{I}_{7a}$ & 0 & $\frac{1}{6}\frac{1}{\ep^2}
                  +\frac{1}{9}\frac{1}{\ep}
                  +\frac{13}{54}+\frac{1}{16} \zeta_2$ \\
           & 1 & $0   $ \\
           & 2 & $ \frac{1}{54} \frac{1}{\ep^2}
                  +\frac{1}{360}\frac{1}{\ep}
                  +\frac{1189}{48600}+\frac{1}{144} \zeta_2  $ \\
             & 3 & $0   $ \\
  \hline
  $\hat{I}_{7b}$ & 0 & $-\frac{2}{15} \frac{1}{\ep^2}
                  +\frac{11}{450} \frac{1}{\ep}
                  -\frac{1643}{13500} - \frac{1}{20} \zeta_2   $ \\
           & 1 & $0   $ \\
           & 2 & $-\frac{2}{189} \frac{1}{\ep^2}
                  +\frac{19}{2205} \frac{1}{\ep}
                  -\frac{225079}{16669800}-\frac{1}{252} \zeta_2   $ \\
           & 3 & $0   $ \\
  \hline
  $\hat{I}_8$    & 0 & $ \frac{1}{6} \frac{1}{\ep^2}
                  +\frac{1}{9} \frac{1}{\ep}
                  +\frac{13}{54} + \frac{1}{16} \zeta_2 $ \\
           & 1 & $ \frac{1}{12} \frac{1}{\ep^2}
                  +\frac{1}{18} \frac{1}{\ep}
                  +\frac{13}{108} +\frac{1}{32} \zeta_2 $ \\
           & 2 & $ \frac{151}{2160} \frac{1}{\ep^2}
                  +\frac{1783}{43200} \frac{1}{\ep}
                  +\frac{785701}{7776000} + \frac{151}{5760} \zeta_2 $ \\
           & 3 & $ \frac{31}{720} \frac{1}{\ep^2}
                  +\frac{1249}{43200} \frac{1}{\ep}
                  +\frac{166801}{2592000} +\frac{31}{1920}  \zeta_2$ \\
  \hline
  $\hat{I}_{9a}$ & 0 & $\frac{1}{\ep^2}
                   +\frac{3}{2} \frac{1}{\ep}
                   +\frac{13}{4} +\frac{3}{8} \zeta_2$ \\
           & 1 & $  \frac{3}{4} \frac{1}{\ep^2}
                   +\frac{59}{48} \frac{1}{\ep}
                   +\frac{1375}{576}+\frac{9}{32} \zeta_2 $ \\       
           & 2 & $  \frac{47}{54} \frac{1}{\ep^2}
                   +\frac{73}{54} \frac{1}{\ep}
                   +\frac{2695}{972} + \frac{47}{144} \zeta_2 $\\
           & 3 & $  \frac{155}{216} \frac{1}{\ep^2}
                   +\frac{2035}{1728} \frac{1}{\ep}
                   +\frac{1424773}{622080}+ \frac{155}{576} \zeta_2 $ \\
  \hline
  $\hat{I}_{9b}$ & 0 & $\frac{1}{\ep^2}
                   +\frac{3}{2} \frac{1}{\ep}
                   +\frac{13}{4} +\frac{3}{8} \zeta_2$ \\
           & 1 & $  \frac{1}{2} \frac{1}{\ep^2}
                   +\frac{13}{24} \frac{1}{\ep}
                   +\frac{413}{288}+\frac{3}{16} \zeta_2 $ \\       
           & 2 & $  \frac{37}{54} \frac{1}{\ep^2}
                   +\frac{101}{108} \frac{1}{\ep}
                   +\frac{8333}{3888}+\frac{37}{144} \zeta_2 $\\
           & 3 & $  \frac{5}{12} \frac{1}{\ep^2} 
                   +\frac{139}{288} \frac{1}{\ep}
                   +\frac{14297}{11520}+\frac{5}{32} \zeta_2 $ \\
  \hline
  \end{tabular}
  \caption{Mellin Moments for Integrals $\hat{I}_{5a}-\hat{I}_{9b}$.}
  \label{TAB2}
  \end{center}
\end{table}

In a first step the different integrals are represented in finite sums over 
hypergeometric series keeping the general 
$\ep$-dependence. After expanding in $\ep$, evanescent poles in summation 
parameters may appear, which have to be dealt with. The resulting indefinite
nested sums can again be calculated by {\sf SIGMA}. For the diagrams  $I_{5a}$ 
and $I_{5b}$ one obtains
\begin{eqnarray}
  I_{5a(b)} &\equiv& \frac{i(\Delta.p)^Na_s^3S_\ep^3}{(m^2)^{2-3\eph}}\hat{I}_{5a(b)}~,
\\
  \hat{I}_{5a} 
  &=& 
   \exp\left(-3\eph\gamma_E\right)
    \Gamma\left(2 - \frac{3}{2} \ep\right)
    B\left(\eph,1+\eph\right) 
    \sum_{j=0}^N 
    \binom{N}{j} 
    \sum_{l=0}^j 
    \binom{j}{l}
    (-1)^{j+l} 
\N\\
&& \frac{B(-\ep+j+2,3-\ep)}{(l+1)(l+2)(N+1-l)} 
   B\left(\ep-1,-\frac{\ep}{2}+j+3\right)
   \left[
   \frac{2}{\ep} 
   + \frac{1}{l + 2 - \frac{\ep}{2}}
+ B\left(-\frac{\ep}{2},l+3\right)\right]
\N\\
  &=& \frac{1}{(N+1) (N+3) (N+4)} \Biggl\{
    4\left[ 
      S_1 
     -\frac{N^2+N-1}{(N+1) (N+2)}
    \right] 
    \frac{1}{\ep^2} 
\N\\ & &
   +\Biggl[ 
      \frac{5}{2} S_1^2 
     -\frac{1}{2} S_2 
     +\frac{-5 N^4 - 18 N^3 + 62 N^2 +289 N + 244}
        {(N+1) (N+2) (N+3) (N+4)} S_1 
\N\\&&
     +\frac{P_4}{(N+1)^2 (N+2)^2 (N+3)(N+4)}
    \Biggr] \frac{1}{\varepsilon} 
\N\\ & &    
   +\Biggl[
      \frac{11}{12} S_1^3
     +\frac{\left(-8 N^4+3 N^3+335 N^2+994 N+736\right)}
        {4(N+1) (N+2) (N+3) (N+4)} S_1^2 \N\\ &&
     +\left(
      \frac{P_5}{2 (N+1)^2 (N+2)^2 (N+3)^2 (N+4)^2}
      +\frac{11}{4} S_2 
      \right) S_1 
\N\\ &&
     +\frac{P_6}{(N+1)^3 (N+2)^3 (N+3)^2 (N+4)^2}
     +\frac{3}{2}\left(
        S_1
       -\frac{\left(N^2+N-1\right)}{(N+1) (N+2)} 
      \right) \zeta_2
\N\\ &&
     +\frac{-2 N^4+9 N^3+185 N^2+580 N+472}{4 (N+1) (N+2) (N+3) (N+4)} S_2
     -\frac{8}{3} S_3 + 6 S_{2,1}
    \Biggr]\Biggr\}+O(\ep),\\
 P_4 &=& -3 N^6 -65 N^5 -415 N^4 -1109 N^3 - 1276 N^2 - 468 N +64,
\\
 P_5 &=& -12 N^8-311 N^7-2943 N^6-13584 N^5-32101 N^4-32407
      N^3+7542 N^2
\N\\ &&
     +40744 N+22784,
\\
P_6 &=& -24 N^9-604 N^8-6089 N^7-32820 N^6-104549 N^5-202546 N^4 
\N\\ &&
     -232976 N^3-143560 N^2-32816 N+3328,
\end{eqnarray}
\begin{eqnarray}
\hat{I}_{5b} &=&  \exp\left(-3\eph\gamma_E\right)
   \Gamma\left(3 - \frac{3}{2} \ep \right)
   \sum_{l=0}^{N} \binom{N}{l} 
   \sum_{j=0}^l \binom{l}{j} 
   \frac{(-1)^{j+l}}{(j+1)(j+2)}
   B\left(\eph, 1+ \eph\right)
\N\\ && 
\times
   B(N+1-j,2) 
   B(l+2-\ep,4-\ep) 
\N\\ && 
\times
   B\left(\ep-1,l+3-\eph\right)
   \left[-\frac{2}{\ep} - B\left(-\eph,j+3\right) 
  -\frac{1}{j+2-\eph}\right]
\\
&=&
  \frac{1}{(N+1)(N+3) (N+4)(N+5)} 
  \Biggl\{
   12 \Biggl[
\frac{
   \left(N^2+N-1\right)}{(N+1)
    (N+2)}- S_1 \Biggr] \frac{1}{\ep^2} \N\\
&& +\Biggl[- 6 S_1^2
+\frac{\left(25
   N^5+261 N^4+775 N^3+3
   N^2-2744 N-2496\right)
   }{(N+1)(N+2)
   (N+3) (N+4)
   (N+5)} S_1 \N\\ &&
+\frac{-N^7+104
   N^6+1497 N^5+7703 N^4+18378
   N^3+20465 N^2+8566
   N+24}{(N+1)^2 (N+2)^2
   (N+3) (N+4) (N+5)} \Biggr]\frac{1}{\ep}
\N\\ &&
+\Biggl[-2 S_1^3
+\frac{\left(10
   N^5+87 N^4+97 N^3-915
   N^2-2699 N-1908\right)
   }{(N+1) (N+2)
   (N+3) (N+4)
   (N+5)} S_1^2
\N
\end{eqnarray}
\begin{eqnarray}
 &&
+\left[\frac{P_7}
{2
   (N+1)^2(N+2)^2 (N+3)^2
   (N+4)^2 (N+5)^2}-12 S_2 \right]
   S_1
\N\\ &&
+\frac{P_8}{2 (N+1)^3 (N+2)^3
   (N+3)^2 (N+4)^2
   (N+5)^2}
\N\\ &&
+ \frac{9}{2} \left[\frac{
   \left(N^2+N-1\right)}{
   (N+1) (N+2) } - 
    S_1 \right] \zeta_2
+\frac{\left(11 N^2-39 N-74\right)
   }{2 (N+1) (N+2)} S_2
+ 5 S_3 - 12
   S_{2,1}\Biggr]\Biggr\}+O(\ep),
\N\\
\\
P_7 &=& 5
   N^{10}+605 N^9+12811
   N^8+124145 N^7+674565
   N^6+2189463 N^5+4196977
   N^4
\N\\ &&
+4214683 N^3+1030490
   N^2-1666304 N-1086816, \\
P_8 &=& -15 N^{12}-497
   N^{11}-5910 N^{10}-27570
   N^9+35363 N^8+1069961
   N^7+5838492 N^6
\N\\ &&
+17154824
   N^5+30447858 N^4+32466210
   N^3+18880180 N^2
\N\\ &&
+4223536
   N-333696. 
\end{eqnarray}
Again both results show a similar structure.

Integral $I_{6}$ is related to $I_{5a}$ by
\begin{eqnarray}
I_6(N) = \left[1+(-1)^N\right] I_{5a}(N+1)~.
\end{eqnarray}

For the diagram $7_{a,b}$ an all-$\ep$ representation without any sums may be obtained. As a 
result only single harmonic sums occur after expanding in $\ep$.
\begin{eqnarray}
I_{7a(b)} &\equiv& \frac{i(\Delta.p)^Na_s^3S_\ep^3}{(m^2)^{2-3\eph}}\hat{I}_{7a(b)},
\\
\hat{I}_{7a} &=& -\exp\left(-3\eph\gamma_E\right)
   \Gamma\left(2 - \frac{3\ep}{2}\right) 
   B\left(\eph,\eph+1\right)
   \frac{[1 + (-1)^N]}{(N+1)(N+2)(N+3)} 
\N\\& & 
   \Biggl\{- \frac{2}{\varepsilon} - \frac{1}{N+3-\eph} - 
   B\left(-\eph, N+4\right) \Biggr\} 
\N\\& & 
   \Gam{-1+\ep, 3+N-\ep, 2-\ep, N+3-\eph}{N+2+\eph,N+5-2 \ep}
\end{eqnarray}
\begin{eqnarray}
   &=& \frac{\left[1 + (-1)^N\right]}{(N+1) (N+3)^2 (N+4) }
   \Biggl\{
    2 \left[S_1 + \frac{2 N +3}{(N+1) (N+2)} \right]
    \frac{1}{\ep^2} 
\N\\& & 
    +\Biggl[\frac{1}{2} \left[S_1^2 + S_2\right]
    -{\frac{ ( 3\,{N}^{4}+18\,{N}^{3}+21\,{N}^{2}-28\,N-40 ) }
           { (N+4)  (N+3) (N+2) (N+1)}} S_1
\N\\ &&
-2\,\frac{ 3\,{N}^{5} +23\,{N}^{4} +53\,{N}^{3} +21\,{N}^{2} -57\,N -48}
         { (N+1)^{2} (N+2)^{2} (N+3) (N+4) } 
    \Biggr]\frac{1}{\ep}
\N\\& & 
   +\Biggl[ \frac{1}{12} S_1^3
    +\frac{1}{6}  S_3
    +{\frac{13}{4}} S_1 S_2
    -\frac{1}{4}
     \frac{ ( 3\,{N}^{4}-6\,{N}^{3}-183\,{N}^{2}-568\,N-472 )}
           { (N+1) (N+2) (N+3) (N+4) } S_2
\N\\& &
    -\frac{ ( 3\,{N}^{4}+18\,{N}^{3}+21\,{N}^{2}-28\,N-40 ) }
          {4 (N+1) (N+2) (N+3) (N+4)} S_1^2
    +\frac{3}{4} \left[
      S_1 
     +\frac{ ( 2\,N+3)}{ (N+1) (N+2) }
    \right] \zeta_2
\N\\& &
   -\frac{P_9}{2 (N+1)^{2} (N+2)^{2} (N+3)^{2} (N+4)^{2}} S_1
\N\\ &&
   -\frac{P_{10}}{2 (N+3)^{2} (N+4)^{2} (N+1)^{3} (N+2)^{3}}
    \Biggr]\Biggr\} +O(\ep)~,
\end{eqnarray}
\begin{eqnarray}
P_9 &=&
36\,{N}^{7}
+511\,{N}^{6}
+2878\,{N}^{5}
+8037\,{N}^{4}
+10942\,{N}^{3}
+4576\,{N}^{2}
\N\\ &&
-4128\,N-3648~, \\
P_{10} &=&
69\,{N}^{8}
+1082\,{N}^{7}
+6983\,{N}^{6}
+23746\,{N}^{5}
+44608\,{N}^{4}
+41876\,{N}^{3}
+7768\,{N}^{2}
\N\\ &&
-17008\,N
-9984~,\\
  \hat{I}_{7b} 
  &=& 
   \exp\left(-3\eph\gamma_E\right)
   \frac{(1+(-1)^N)}{(N+1) (N+2) (N+3)} 
   \Gamma\left(3-\frac{3}{2} \ep \right) 
   B\left(3-\ep,N+3-\ep\right) 
\N\\ && 
   \times
   B\left(\eph,1+\eph\right) 
   B\left(\ep-1,N+3-\eph\right) 
   \left[
    -B\left(-\eph,N+4\right)
    -\frac{1}{N+3-\eph} 
    -\frac{2}{\ep}
   \right]
\N
\\
  &=& 
   \frac{[1+(-1)^N]}{(N+1) (N+3)^2 (N+4) (N+5)} 
   \Biggl\{
    -8 \Biggl[  
     S_1 +
     \frac{\left( 3+2\,N \right)}{(N+1) (N+2)}
     \Biggr] \frac{1}{\ep^2}
\N\\ & &
    +2\Biggl[
    -\left(S_1^2 +S_2\right)
    +\frac {11\,{N}^{5}+133\,{N}^{4}+567\,{N}^{3}+999\,{N}^{2}+610\,N+8}
       {(N+1)  (N+2) (N+3) (N+4) (N+5)} S_1
\N\\
&& +
     \frac { 22\,{N}^{6} +301\,{N}^{5} +1563\,{N}^{4} +3869\,{N}^{3} +4667\,{N}^{2} +2394\,N +264}
       {(N+1)^2 (N+2)^2 (N+3) (N+4) (N+5)} \Biggr] \frac{1}{\ep}
\N\\ &&
    +\Biggl[ 
      -\frac{2}{3} S_3 
      -\frac{1}{3} S_1^3 
      -13 S_1 S_2 
      -3 S_1 \zeta_2 
      -3\,\frac{ \left( 3+2\,N \right)}{(N+1)(N+2)} \zeta_2
\N\\
&&
      +\frac { 11\,{N}^{5}+133\,{N }^{4}+567\,{N}^{3}+999\,{N}^{2}+610\,N+8 }
         {2 (N+1) (N+2) (N+3) (N+4) (N+5)} S_1^2
     \Biggr]
\N\\ &&
    +\frac {11\,{N}^{5}+85\,{N}^{4}-81\,{N}^{3}-2121\,{N}^{2}-5654\,N-4312 }
       {2(N+1) (N+2) (N+3) (N+4) (N+5)} S_2
\N
\\ 
&&
    -2\,\frac { P_{11}}{(N+1)^{2} (N+2)^{2} (N+3)^{2} (N+4)^{2} (N+5)^{2}} S_1
\N
\end{eqnarray}
\begin{eqnarray}
&&
    -\frac {P_{12}}{ (N+1)^{3} (N+2)^{3} (N+3)^{2} (N+4)^{2} (N+5)^{2}}
   \Biggr\} +O(\ep),
\\
P_{11} &=& 
9\,{N}^{10}
+182\,{N}^{9}
+1388\,{N}^{8}
+4103\,{N}^{7}
-4913\,{N}^{6}
-72860\,{N}^{5}
-225446\,{N}^{4}
\N\\ &&
-327313\,{N}^{3}
-198070\,{N}^{2}
+17240\,N
+52416, 
\\
P_{12} &=& 
36\,{N}^{11}
+793\,{N}^{10}
+6942\,{N}^{9}
+28237\,{N}^{8}
+28250\,{N}^{7}
-224189\,{N}^{6}
\N\\ && 
-1079534\,{N}^{5}
-2213865\,{N}^{4}
-2276462\,{N}^{3}
-830640\,{N}^{2}
\N\\ &&
+388496\,N
+315456.
\end{eqnarray}

Diagram~8, despite being expressed only by a threefold sum, is more demanding. One obtains
\begin{eqnarray}
  I_{8} &\equiv& \frac{i(\Delta.p)^Na_s^3S_\ep^3}{(m^2)^{2-3\eph}}\hat{I}_{8},
\\
  \hat{I}_8 
  &=& 
   -\exp\left(-3\eph\gamma_E\right) 
    \Gamma(\ep-1) 
    \Gamma\left(2-\frac{3}{2} \ep\right) 
    B\left(\eph,1+\eph\right)
    \sum_{i=0}^N 
    \sum_{j=0}^{N-i} 
    \binom{N-i}{j} 
    (-1)^j 
\N\\
 && 
    \times
    \sum_{l=0}^{i+j} 
    \binom{i+j}{l} 
    (-1)^l 
    \Gam{3+j+i-\eph,3+i-\ep,2+j-\ep}{5+i+j-2 \ep,2+i+j+\eph}
\N\\ && 
    \times 
    \frac{1}{(l+1) (l+2) (N+1-l)}
    \left[B\left(-\eph,1\right)-B\left(l+2-\eph,1\right)
   -B\left(-\eph,l+3\right)\right] 
\\
  &=& 
    \frac{1}{(N+2) (N+4) (N+5)}\Biggl\{\Biggl[
    \Biggl(
      \frac{2 (-1)^N \left(N^2+5 N+7\right)}{(N+2) (N+3)^2}
\N\\& &
     +\frac{2 \left(2 N^3+13 N^2+27 N+20\right)}{(N+1) (N+3)^2}
    \Biggr) S_{1}
         +S_{1}^2
         +3 S_{2}
         +\frac{2 (-1)^N (2 N^3+13 N^2+29 N+21)}{(N+1) (N+2)^2 (N+3)^2}
\N\\& &
         -\frac{2 \left(2 N^6+18 N^5+57 N^4+60 N^3-53 N^2-163 N-99\right)}
               {(N+1)^2 (N+2)^2 (N+3)^2}
         \Biggr]\frac{1}{\ep^2}
\N\\& &
         +\frac{1}{N+3}\Biggl[
          \frac{1}{2} (N+3) S_{1}^3
         +\Biggl(\frac{(-1)^N \left(N^2+5 N+7\right)}{2 (N+2) (N+3)}
\N\\& &
         +\frac{2 N^6+43 N^5+360 N^4+1529 N^3+3524 N^2+4218 N+2048}
               {2 (N+1) (N+2) (N+3) (N+4) (N+5)}\Biggr) S_{1}^2
\N\\& &
         +\Biggl(
          \frac{P_{13}}{(N+1)^2 (N+2) (N+3)^2 (N+4) (N+5)}
\N\\& &
         +\frac{(-1)^N P_{14}}{(N+1)^2 (N+2)^2 (N+3)^2 (N+4) (N+5)}
\N\\& &
         +4 S_{-2}
          \Biggr) S_{1}
         +\Biggl(
          \frac{7}{2}(N+3) S_{1}
         +\frac{(-1)^N \left(N^2+5 N+7\right)}{2 (N+2) (N+3)}
\N\\& &
         +\frac{-10 N^6-133 N^5-612 N^4-915 N^3+1052 N^2+4246 N+3104}{2 (N+1) (N+2) (N+3) (N+4) (N+5)}\Biggr) S_{2}
\N
\\
& &
         +\frac{4 (2 N+3)}{(N+1) (N+2)} S_{-2}
         +2(N+5) S_{3}
         -4(N+3) S_{2,1}
\N
\end{eqnarray}
\begin{eqnarray}
& &
         +\frac{(-1)^N P_{15}}{(N+1)^3 (N+2)^3 (N+3)^2 (N+4) (N+5)}
\N\\
& &
         +\frac{P_{16}}{(N+1)^3 (N+2)^3 (N+3)^2 (N+4) (N+5)}
         \Biggr] \frac{1}{\ep} 
\N\\
& &
         + \frac{1}{N+3}\Biggl[
           \frac{7}{48} (N+3) S_{1}^4
         +\Biggl(
          \frac{(-1)^N \left(N^2+5 N+7\right)}{12 (N+2) (N+3)}
\N\\& &
         +\frac{2 N^6+59 N^5+588 N^4+2805 N^3+7040 N^2+8974 N+4544}
               {12 (N+1) (N+2) (N+3) (N+4) (N+5)}
          \Biggr) S_{1}^3
\N\\& &
         +\Biggl(\frac{(-1)^N P_{17}}{4 (N+1)^2 (N+2)^2 (N+3)^2 (N+4) (N+5)}
\N\\& &
         +\frac{P_{18}}{4 (N+1)^2 (N+2)^2 (N+3)^2 (N+4)^2 (N+5)^2}
         +7 S_{-2}\Biggr) S_{1}^2
\N\\
 & &
         +\Biggl(\frac{(-1)^N P_{19}}{2 (N+1)^3 (N+2)^3 (N+3)^3 (N+4)^2 (N+5)^2}
\N\\& &
         +\frac{P_{20}}{2 (N+1)^3 (N+2)^2 (N+3)^3 (N+4)^2 (N+5)^2}
         +5 S_{-3}
\N\\& &
         -\frac{2 \left(5 N^5+49 N^4+104 N^3-285 N^2-1213 N-1036\right) S_{-2}}{(N+1) (N+2) (N+3) (N+4) (N+5)}\Biggr) S_{1}
         +\frac{(55 N+141)}{16} S_{2}^2
\N\\& &
         +\frac{(-1)^N P_{21}}{2 (N+1)^4 (N+2)^4 (N+3)^3 (N+4)^2 (N+5)^2}
\N\\& &
         +\frac{P_{22}}{2 (N+1)^4 (N+2)^4 (N+3)^3 (N+4)^2 (N+5)^2}
         +\frac{5 (2 N+3)}{(N+1) (N+2)} S_{-3}
\N\\& &
         -\frac{4 \left(5 N^6+63 N^5+275 N^4+425 N^3-160 N^2-1004 N-684\right) S_{-2}}{(N+1)^2 (N+2)^2 (N+3) (N+4) (N+5)}
\N\\& &
         +\Biggl(\frac{3 (9 N+31)}{8} S_{1}^2
         +\Biggl(\frac{13 (-1)^N \left(N^2+5 N+7\right)}{4 (N+2) (N+3)}
\N\\& &
         +\frac{-10 N^6-65 N^5+420 N^4+5213 N^3+18860 N^2+29514 N+16976}{4 (N+1) (N+2) (N+3) (N+4) (N+5)}\Biggr) S_{1}
\N\\& &
         +\frac{(-1)^N P_{23}}{4 (N+1)^2 (N+2)^2 (N+3)^2 (N+4) (N+5)}
\N\\& &
         +\frac{P_{24}}{4 (N+1)^2 (N+2)^2 (N+3)^2 (N+4)^2 (N+5)^2}
         +S_{-2}\Biggr) S_{2}
         +\zeta_2 \Biggl(\frac{3}{8} (N+3) S_{1}^2
\N\\& &
         +\Biggl(\frac{3 (-1)^N \left(N^2+5 N+7\right)}{4 (N+2) (N+3)}
         +\frac{3 \left(2 N^3+13 N^2+27 N+20\right)}{4 (N+1) (N+3)}\Biggr) S_{1}
\N\\& &
         -\frac{3 \left(2 N^6+18 N^5+57 N^4+60 N^3-53 N^2-163 N-99\right)}{4 (N+1)^2 (N+2)^2 (N+3)}
\N\\& &
         +\frac{3 (-1)^N \left(2 N^3+13 N^2+29 N+21\right)}{4 (N+1) (N+2)^2 (N+3)}
         +\frac{9}{8}(N+3) S_{2}\Biggr)
         +\Biggl(\frac{(-1)^N \left(N^2+5 N+7\right)}{6 (N+2) (N+3)}
\N
\end{eqnarray}
\begin{eqnarray}
& &
         +\frac{-34 N^5-383 N^4-1379 N^3-1280 N^2+1830 N+2632}{6 (N+1) (N+2) (N+3) (N+4)}
         +\frac{(13 N+105)}{6} S_{1}\Biggr) S_{3}
\N\\& &
         +\frac{(53-N)}{8} S_{4}
         +\left(-\frac{6 (2 N+3)}{(N+1) (N+2)}
         -6 S_{1}\right) S_{-2,1}
\N\\& &
         +\left(\frac{12 N^5+140 N^4+546 N^3+725 N^2-93 N-532}{(N+1) (N+2) (N+4) (N+5)}
         +(-4 N-15) S_{1}\right) S_{2,1}
\N\\& &
         +(N-11) S_{3,1}
         +(N+9) S_{2,1,1}\Biggr]\Biggr\} + O(\ep)~,
\\
P_{13}&=&-5 N^8-68 N^7-264 N^6+410 N^5+6293 N^4+20720 N^3+32900 N^2
\N\\& &
+26206 N+8440,\\
P_{14}&=&-3 N^8-49 N^7-321 N^6-1069 N^5-1863 N^4-1559 N^3-773 N^2
\N\\ &&
-1199 N-1108,\\
P_{15}&=&-6 N^9-108 N^8-810 N^7-3288 N^6-7855 N^5-11456 N^4-11282 N^3
\N\\& &
-10300 N^2-9171 N-4164,\\
P_{16}&=&-3 N^{11}-112 N^{10}-1610 N^9-12443 N^8-58690 N^7-178509 N^6-355289 N^5
\N\\& &
-451853 N^4-334491 N^3-98371 N^2+31775 N+23364,\\
P_{17}&=&-3 N^8-49 N^7-321 N^6-1069 N^5-1863 N^4-1559 N^3-773 N^2
\N\\ &&
-1199 N-1108,\\
P_{18}&=&-8 N^{11}-189 N^{10}-1643 N^9-4234 N^8+32416 N^7+340621 N^6+1490447 N^5
\N\\& &
+3864842 N^4+6329756 N^3+6460920 N^2+3775088 N+971008,\\
P_{19}&=&-42 N^{12}-1213 N^{11}-15525 N^{10}-115864 N^9-557609 N^8-1804421 N^7-3966084 
N^6
\N\\& &
-5845058 N^5-5625111 N^4-3597908 N^3-2035597 N^2-1373344 N-553968,\\
P_{20}&=&-12 N^{13}-523 N^{12}-9558 N^{11}-98647 N^{10}-644321 N^9-2799010 N^8-8183392 
N^7
\N\\& &
-15639871 N^6-17214281 N^5-4125073 N^4+17049900 N^3+25968164 N^2
\N\\& &
+16422416 N+4131840,\\
P_{21}&=&-81 N^{13}-2458 N^{12}-33378 N^{11}-267579 N^{10}-1405780 N^9-5075289 N^8
\N\\& &
-12828559 N^7-22692458 N^6-27711081 N^5-23127963 N^4-14102081 N^3
\N\\& &
-8182893 N^2-4780496 N-1528944,\\
P_{22}&=&-60 N^{15}-2640 N^{14}-51484 N^{13}-594504 N^{12}-4564031 N^{11}-24724313 
N^{10}
\N\\& &
-97683496 N^9-286337829 N^8-626024531 N^7-1014709686 N^6-1194939874 N^5
\N\\& &
-978463105 N^4-504961532 N^3-120080691 N^2+14776800 N+11512944,\\
P_{23}&=&-3 N^8-25 N^7+147 N^6+2723 N^5+14685 N^4+40381 N^3+60691 N^2
\N\\& &
+46645 N+14012,\\
P_{24}&=&-2 N^{11}-177 N^{10}-3713 N^9-36850 N^8-204686 N^7-647555 N^6-952035 N^5
\N\\& &
+618266 N^4+5332620 N^3+9769044 N^2+8340336 N+2862784.
\end{eqnarray}
The contributions to diagram~$9a$ are derived via sixfold nested sums:
\begin{eqnarray}
I_{9a/b} &\equiv&
 \frac{i(\Delta.p)^Na_s^3S_\ep^3}{(m^2)^{1-3\eph}}\hat{I}_{9a/b},
\end{eqnarray}
\begin{eqnarray}
\hat{I}_{9a} &=&  \exp\left(-3\eph\gamma_E\right)
   \Gamma\left(1-3\eph\right)
   \sum_{j=0}^{N}\binom{N+2}{j+2}
   \sum_{k=0}^{j}\binom{j+1}{k+1}
   \sum_{l=0}^{k}\binom{k}{l}  
   (-1)^{k+l}
\N\\
&&
   \sum_{q=0}^{N-j}\binom{N-j}{q}
   (-1)^{N-j-q}
   \sum_{r_2=0}^{N-l-q}\binom{N-l-q}{r_2}
\N\\&&
   \sum_{r_1=0}^{N-l-q-r_2}\binom{N-l-q-r_2}{r_1}
   \frac{B(1-\ep,N+2-j-\ep)
   B(\eph,k+1+\eph)}{(N+1-q-r_1-r_2)(q+r_2+1)}
\N\\&&
   B(r_2+\ep,r_1+1)
   B\left(N+1-l-q-r_1-r_2-\eph,r_1+r_2+1+\ep\right) 
\N\\
&=&
   \frac{1}{(N+3) (N+4)}
   \Biggl\{
     \Biggl[
   -\frac{4 \left(N^3+3 N^2-N-5\right)}{(N+1) (N+2) (N+3)} S_{1}
   +2 S_{1}^2
   +\frac{4 (-1)^N}{N+3} S_{1}
   +4 S_{-2}
\N\\&&
   +2 (2 N+5) S_{2}
   +\frac{4 (-1)^N \left(2 N^3+7 N^2+4 N-3\right)}{(N+1)^2 (N+2)^2 (N+3)}
   +\frac{4 \left(6 N^3+34 N^2+63 N+39\right)}{(N+1)^2 (N+2)^2 (N+3)}
      \Biggr]
      \frac{1}{\ep^2}
\N\\&&
     +\Biggl[
    \frac{\left(-4 N^4-25 N^3-30 N^2+49 N+76\right)}{(N+1) (N+2) (N+3) (N+4)} S_{1}^2
   -\frac{4 \left(2 N^4+14 N^3+27 N^2+5 N-16\right)}{(N+1) (N+2) (N+3) (N+4)} S_{-2}
\N\\
&&
   +\frac{\left(10 N^4+73 N^3+158 N^2+73 N-52\right)}{(N+1) (N+2) (N+3) (N+4)} S_{2}
\N\\&&
   +\frac{2 (-1)^N \left(12 N^5+127 N^4+538 N^3+1177 N^2+1354 N+648\right)}{(N+1)^2 (N+2)^2 (N+3)^2 (N+4)} S_{1}
\N\\&&
   -\frac{2 \left(8 N^6+51 N^5-72 N^4-1330 N^3-4062 N^2-5151 N-2436\right)}{(N+1)^2 (N+2)^2 (N+3)^2 (N+4)} S_{1}
\N\\&&
   +S_{1}^3
   +\frac{(-1)^N}{N+3}\left(
      S_{1}^2
     -S_{2}
   \right)
   +4 S_{-2} S_{1}
   -5 S_{2} S_{1}
   +2 (4 N+15) S_{-3}
   +2 (N-1) S_{3}
\N\\&&
   -12 S_{-2,1}
   +8 (N+4) S_{2,1}
\N\\&&
   +\frac{2 (-1)^N \left(11 N^6+60 N^5-160 N^4-1837 N^3-5005 N^2-5801 N-2508\right)}{(N+1)^3 (N+2)^3 (N+3)^2 (N+4)}
\N\\&&
   +\frac{2 \left(70 N^6+893 N^5+4640 N^4+12626 N^3+19074 N^2+15269 N+5100\right)}{(N+1)^3 (N+2)^3 (N+3)^2 (N+4)}
     \Biggr]
     \frac{1}{\ep}
\N\\& &
   +\frac{7}{24} S_{1}^4
   +\frac{\left(-10 N^4-61 N^3-68 N^2+129 N+188\right)}{6 (N+1) (N+2) (N+3) (N+4)} S_{1}^3
\N\\&&
   +\frac{(-1)^N \left(12 N^5+127 N^4+538 N^3+1177 N^2+1354 N+648\right)}{2 (N+1)^2 (N+2)^2 (N+3)^2 (N+4)} S_{1}^2
\N\\&&
   +\frac{P_{25}}{2 (N+1)^2 (N+2)^2 (N+3)^2 (N+4)^2} S_{1}^2
   +\frac{3}{4} \zeta_2 S_{1}^2
   -4 S_{-2} S_{1}^2
   -\frac{13}{4} S_{2} S_{1}^2
\N\\&&
   +\frac{(-1)^N P_{26}}{(N+1)^3 (N+2)^3 (N+3)^3 (N+4)^2} S_{1}
   +\frac{P_{27}}{(N+1)^3 (N+2)^3 (N+3)^3 (N+4)^2} S_{1}
\N\\&&
   -\frac{3 \left(N^3+3 N^2-N-5\right)}{2 (N+1) (N+2) (N+3)} \zeta_2 S_{1}
   -2 S_{-3} S_{1}
\N\\&&
   -\frac{4 \left(4 N^4+41 N^3+155 N^2+254 N+148\right)}{(N+1) (N+2) (N+3) (N+4)} S_{-2} S_{1}
\N\\ &&
   +\frac{(-1)^N}{N+3}\left(
   -4 S_{-2} S_{1}
   +\frac{9}{2} S_{2} S_{1}
   +\frac{3}{2} \zeta_2 S_{1}
   +\frac{1}{6} S_{1}^3
   -2 S_{-3}
   +\frac{10}{3} S_{3}
   +2 S_{2,1}
   +12 S_{-2,1}
   \right)
\N
\end{eqnarray}
\begin{eqnarray}
&&
   +\frac{\left(-14 N^4-201 N^3-936 N^2-1715 N-1044\right)}{2 (N+1) (N+2) (N+3) (N+4)} S_{2} S_{1}
   -\frac{119}{3} S_{3} S_{1}
\N\\&&
   -12 S_{-2,1} S_{1}
   +22 S_{2,1} S_{1}
   -2 S_{-2}^2
   +\frac{1}{8} (32 N+119) S_{2}^2
\N\\&&
   +\frac{(-1)^N P_{28}}{(N+1)^4 (N+2)^4 (N+3)^3 (N+4)^2}
   +\frac{P_{29}}{(N+1)^4 (N+2)^4 (N+3)^3 (N+4)^2}
\N\\&&
   +\frac{3 (-1)^N \left(2 N^3+7 N^2+4 N-3\right)}{2 (N+1)^2 (N+2)^2 (N+3)} \zeta_2
   +\frac{3 \left(6 N^3+34 N^2+63 N+39\right)}{2 (N+1)^2 (N+2)^2 (N+3)} \zeta_2
\N\\&&
   +(8 N+39) S_{-4}
   +\frac{2 \left(8 N^5+108 N^4+558 N^3+1365 N^2+1553 N+640\right)}{(N+1) (N+2) (N+3) (N+4)} S_{-3}
\N\\&&
   -\frac{4 (-1)^N \left(2 N^3+7 N^2+4 N-3\right)}{(N+1)^2 (N+2)^2 (N+3)} S_{-2}
   -\frac{4 P_{30}}{(N+1)^2 (N+2)^2 (N+3)^2 (N+4)^2} S_{-2}
\N\\&&
   +\frac{3}{2} \zeta_2 S_{-2}
   +\frac{(-1)^N \left(8 N^5+79 N^4+186 N^3-279 N^2-1426 N-1224\right)}{2 (N+1)^2 (N+2)^2 (N+3)^2 (N+4)} S_{2}
\N\\&&
   +\frac{P_{31}}{2 (N+1)^2 (N+2)^2 (N+3)^2 (N+4)^2} S_{2}
   +\frac{3}{4} (2 N+5) \zeta_2 S_{2}
   +8 S_{-2} S_{2}
\N
\end{eqnarray}
\begin{eqnarray}
&&
   +\frac{\left(-18 N^5-229 N^4-1498 N^3-5558 N^2-10017 N-6460\right)}{3 (N+1) (N+2) (N+3) (N+4)} S_{3}
\N\\&&
   +\frac{1}{4} (20 N-29) S_{4}
   -14 S_{-3,1}
   +\frac{4 \left(4 N^4+22 N^3+11 N^2-85 N-96\right)}{(N+1) (N+2) (N+3) (N+4)} S_{-2,1}
\N\\&&
   -14 S_{-2,2}
   +\frac{2 \left(11 N^4+107 N^3+397 N^2+640 N+361\right)}{(N+1) (N+2) (N+3)} S_{2,1}
   +2 (N+36) S_{3,1}
\N\\&&
   +28 S_{-2,1,1}
   +2 (2 N-7) S_{2,1,1}\Biggr\}
  + O(\ep)~,
\\
P_{25} &=& -6 N^8-164 N^7-1613 N^6-7762 N^5-19526 N^4-22888 N^3-2137 N^2
\N\\&&
+19968 N+13264~,\\
P_{26} &=& 119 N^8+2250 N^7+18755 N^6+90365 N^5+275464 N^4+542281 N^3+668958 N^2
\N\\&&
+469072 N+142112~,\\
P_{27} &=& 16 N^{11}+448 N^{10}+5568 N^9+41171 N^8+204092 N^7+720291 N^6+1858328 N^5
\N\\&&
+3504939 N^4+4712624 N^3+4272331 N^2+2335952 N+581072~,\\
P_{28} &=& 78 N^9+937 N^8+2466 N^7-17638 N^6-155141 N^5-538674 N^4-1047495 N^3
\N\\&&
-1197445 N^2-757472 N-206256~,\\
P_{29} &=& 568 N^9+11297 N^8+98332 N^7+492027 N^6+1561688 N^5+3266831 N^4
\N\\&&
+4516420 N^3+3994885 N^2+2061840 N+475824~,\\
P_{30} &=& 4 N^8+96 N^7+942 N^6+4995 N^5+15753 N^4+30351 N^3+34903 N^2
\N\\&&
+21844 N+5648~,\\
P_{31} &=& -32 N^9-730 N^8-7180 N^7-40057 N^6-139918 N^5-317434 N^4-466820 N^3
\N\\&&
-426421 N^2-216416 N-45040~,
\end{eqnarray}
For diagram $9b$ one obtains
\begin{eqnarray}
\hat{I}_{9b} &=& 
   \exp\left(-3\eph\gamma_E\right)\Gamma\left(1 - \frac{3}{2}\ep\right) 
   \sum_{j=0}^{N}   \binom{N+2}{j+2} 
   \sum_{k=0}^j     \binom{j+1}{k+1} 
   \sum_{l=0}^k     \binom{k}{l} 
\N
\end{eqnarray}
\begin{eqnarray}
&&
\times 
   \sum_{q=0}^{N-j} \binom{N-j}{q} 
   (-1)^{N-j-q+k} 
   \sum_{r_1=0}^{N-l-q} \binom{N-l-q}{r_1} 
   \sum_{r_2=0}^{N-l-q-r_1} \binom{N-l-q-r_1}{r_2} 
\N\\ && 
\times
   B(1-\ep,N+2-j-\ep)
   B\left(k-l+\eph,l+1+\eph\right)
   B(r_2+\ep,r_1+1) 
\N\\ && 
\times
   B\left(N+1-l-q-r_1-r_2-\eph, r_1+r_2+1+\ep\right)
   \frac{1}{(q+r_2+1)(N+1-q-r_1-r_2)} 
\N\\ 
&=&
   \frac{1}{(N+3) (N+4)} 
   \Biggl\{
    \Biggl[
    4 S_{1}^2
   +\frac{8 (2 N+3)}{(N+1) (N+2)} S_{1}
   -4 (-1)^N S_{-2}
   +\frac{8 (2 N+3)}{(N+1)^2 (N+2)}
   \Biggr]\frac{1}{\ep^2}
\N\\ &&
   +\Biggl[
    \frac{2 (-1)^N (2 N^2+14 N+17)}{(N+1)^2 (N+2)^2} S_{1}
   +\frac{2 (14 N^3+103 N^2+235 N+164)}{(N+1) (N+2) (N+3) (N+4)} S_{1}^2
\N\\ &&
   -\frac{4 (-1)^N (6 N^3+43 N^2+95 N+64)}{(N+1) (N+2) (N+3) (N+4)} S_{-2}
\N\\ &&
   +\frac{2 (44 N^4+376 N^3+1135 N^2+1445 N+660)}{(N+1)^2 (N+2)^2 (N+3) (N+4)} S_{1}
\N\\ &&
   +2 S_{1}^3
   -4 (-1)^N S_{-2} S_{1}
   +8 S_{-2} S_{1}
   -2 (-1)^N S_{2} S_{1}
   -2 (-1)^N S_{-3}
   +8 S_{-3}
   +4 (-1)^N S_{2,1}
\N\\
 &&
   +\frac{2(2 N+3)}{(N+1) (N+2)}\Bigl(
    4 S_{-2}
   - (-1)^N S_{2}
   \Bigr)
   -2 (-1)^N S_{3}
   +4 (-1)^N S_{-2,1}
   -16 S_{-2,1}
\N\\ &&
   +\frac{2 (42 N^4+355 N^3+1056 N^2+1319 N+588)}{(N+1)^3 (N+2)^2 (N+3) (N+4)}
   -\frac{2 (-1)^N (8 N+13)}{(N+1)^2 (N+2)^2}
   \Biggr]
   \frac{1}{\ep}
\N\\ &&
   +\frac{7}{12} S_{1}^4
   +\frac{\left(38 N^3+275 N^2+615 N+420\right)}{3 (N+1) (N+2) (N+3) (N+4)} S_{1}^3
   +\frac{(-1)^N \left(2 N^2+14 N+17\right)}{2 (N+1)^2 (N+2)^2} S_{1}^2
\N\\ &&
   +\frac{\left(260 N^6+3844 N^5+23111 N^4+72230 N^3+123747 N^2+110376 N+40304\right)}{2 (N+1)^2 (N+2)^2 (N+3)^2 (N+4)^2} S_{1}^2
   +\frac{3}{2} \zeta_2 S_{1}^2
\N\\ &&
   +10 S_{-2} S_{1}^2
   +\frac{13}{2} S_{2} S_{1}^2
   +\frac{(-1)^N \left(-3 N^5+8 N^4+197 N^3+631 N^2+647 N+148\right)}{(N+1)^3 (N+2)^3 (N+3) (N+4)} S_{1}
\N\\ &&
   +\frac{P_{32}}{(N+1)^3 (N+2)^3 (N+3)^2 (N+4)^2} S_{1}
   +\frac{4 \left(18 N^3+137 N^2+325 N+236\right)}{(N+1) (N+2) (N+3) (N+4)} S_{-2} S_{1}
\N\\ &&
   +\frac{(-1)^N (2 N+7)}{(N+3) (N+4)}\left(
    -8 S_{-2} S_{1}
    -4 S_{2} S_{1}
    +8 S_{-2,1}
   \right)
   -4 (-1)^N S_{-3} S_{1}
   +26 S_{-3} S_{1}
\N\\ &&
   +\frac{(2 N+3)}{(N+1) (N+2)}\left(
     3 \zeta_2 S_{1}
    +13 S_{2} S_{1}
    +\frac{29}{3} S_{3}
    +2 S_{2,1}
   \right)
   +\frac{29}{3} S_{3} S_{1}
   -28 S_{-2,1} S_{1}
\N\\ &&
   +2 S_{2,1} S_{1}
   +(-1)^N S_{3} S_{1}
   -2 (-1)^N S_{2,1} S_{1}
   +2 (-1)^N S_{-2}^2
   -(-1)^N S_{2}^2
   +\frac{3}{4} S_{2}^2
\N\\ &&
   +\frac{(-1)^N \left(-127 N^6-1569 N^5-7862 N^4-20557 N^3-29733 N^2-22680 N-7168\right)}{(N+1)^4 (N+2)^4 (N+3) (N+4)}
\N\\ &&
   +\frac{P_{33}}{(N+1)^4 (N+2)^4 (N+3)^2 (N+4)^2}
   +\frac{3 (2 N+3)}{(N+1)^2 (N+2)} \zeta_2
   -7 (-1)^N S_{-4}
   +14 S_{-4}
\N\\ &&
   -\frac{8 (-1)^N \left(2 N^3+15 N^2+35 N+25\right)}{(N+1) (N+2) (N+3) (N+4)} S_{-3}
   +\frac{2 \left(42 N^3+325 N^2+785 N+580\right)}{(N+1) (N+2) (N+3) (N+4)} S_{-3}
\N\\ &&
   +\frac{4 \left(25 N^4+216 N^3+656 N^2+831 N+372\right)}{(N+1)^2 (N+2)^2 (N+3) (N+4)} S_{-2}
\N
\end{eqnarray}
\begin{eqnarray}
&&
   -\frac{16 (-1)^N \left(6 N^6+86 N^5+500 N^4+1508 N^3+2491 N^2+2145 N+760\right)}{(N+1)^2 (N+2)^2 (N+3)^2 (N+4)^2} S_{-2}
\N\\ &&
   -\frac{3}{2} (-1)^N S_{-2} \zeta_2
   +\frac{\left(64 N^2+216 N+183\right)}{2 (N+1)^2 (N+2)^2} S_{2}
\N\\ &&
   +\frac{(-1)^N \left(-30 N^4-212 N^3-437 N^2-177 N+156\right)}{2 (N+1)^2 (N+2)^2 (N+3) (N+4)} S_{2}
   -8 (-1)^N S_{-2} S_{2}
\N\\ &&
   -2 S_{-2} S_{2}
   +\frac{(-1)^N \left(-6 N^3-35 N^2-55 N-20\right)}{(N+1) (N+2) (N+3) (N+4)} S_{3}
\N\\ &&
   +2 (-1)^N S_{4}
   +\frac{19}{2} S_{4}
   +4 (-1)^N S_{-3,1}
   -24 S_{-3,1}
\N\\ &&
   -\frac{4 \left(30 N^3+223 N^2+515 N+364\right)}{(N+1) (N+2) (N+3) (N+4)} S_{-2,1}
   +2 (-1)^N S_{-2,2}
   -20 S_{-2,2}
\N\\ &&
   +\frac{2 (-1)^N \left(6 N^3+35 N^2+55 N+20\right)}{(N+1) (N+2) (N+3) (N+4)} S_{2,1}
   -13 (-1)^N S_{3,1}
   -17 S_{3,1}
\N\\ &&
   +32 S_{-2,1,1}
   +5 (-1)^N S_{2,1,1}
   +3 S_{2,1,1}\Biggr\}
   +O(\ep)~,
\end{eqnarray}
with
\begin{eqnarray}
P_{32}&=&367 N^7+5827 N^6+38741 N^5+139834 N^4+296246 N^3+369049 N^2
\N\\&&
+251056 N+72240~,\\
P_{33}&=&373 N^8+6728 N^7+52275 N^6+228755 N^5+617580 N^4+1055293 N^3
\N\\&&
+1117044 N^2+671360 N+175872~.
\end{eqnarray}

The diagrams yield the harmonic sums of maximum depth 3 and maximum weight 4:
\begin{eqnarray}
  S_{1},\;
  S_{2},\;
  S_{3},\;
  S_{4},\;
  S_{-2},\;
  S_{-3},\;
  S_{-4},\;
  S_{2,1},\;
  S_{-2,1},\;
  S_{-2,2},\;
  S_{3,1},\;
  S_{-3,1},\;
  S_{2,1,1},\;
  S_{-2,1,1}.
\end{eqnarray}
This set is the same as for the $O(\alpha_s^2\ep)$ contributions to the massive
OMEs, which contribute at the three loop order via renormalization
\cite{BBK4} and for a wide class of other processes, see \cite{BR}.  In 
addition, in the case of the diagrams with six massive
propagators also the following generalized harmonic sums \cite{MUW,ABS1} contribute:
\begin{eqnarray}
  S_1\Bigl(2\Bigr),
  S_{1,1,1}\left(\frac{1}{2},1,1\right),
  S_{1,2}\left(\frac{1}{2},1\right),
  S_{1,1,2}\left(2,\frac{1}{2},1\right),
  S_{1,1,1,1}\left(2,\frac{1}{2},1,1\right),
\end{eqnarray}
However, these terms do not contribute to the diagrams with three
massive propagators. 

Representations for the analytic continuation of the harmonic sums to $N \in 
\mathbb{C}$ were calculated in \cite{HBAS2,ANCONT} and the inverse Mellin 
transforms of the generalized harmonic sums are given in (\ref{eqLMT1}--\ref{eqLMT2}).

\section{Functions from Moments}
\label{sec:4}
\renewcommand{\theequation}{\thesection.\arabic{equation}}
\setcounter{equation}{0} 

\vspace{1mm}
\noindent
Since the Feynman integrals $I_1 - I_{9b}$ are recurrent quantities in $N$
one may in principle find their analytic form for general values of $N$
using the method described in Ref.~\cite{KAUERS}. In the following we determine 
the amount of Mellin moments needed to find these solutions and solve the recurrences
being obtained by {\sf Sigma}~\cite{SIGMA}. This method has been used before in 
Ref.~\cite{LINEQ} in case of very large recurrences.

In Tables~3 and 4 we
summarize the results  for diagrams $I_{1a}$ to $I_{4}$ and $I_{5a}$ to
$I_{9b}$, respectively. Here, we ignored pre-factors of the form $[1+(-1)^N]$
which can always be identified in the analytic calculation and used the
remainder part to evaluate the moments. In general, this method is more
efficient since lower moments have to be computed to establish the
corresponding recurrences.

\begin{table}[htbp]
  \begin{center}
  \renewcommand{\arraystretch}{1.3} 
  \begin{tabular}{|l||r|r|r||r|r|r|}
\hline
\multicolumn{1}{|c||}{ } &
\multicolumn{3}{|c||}{rational} &
\multicolumn{3}{|c|}{$\zeta_3$} \\
\cline{2-7}
\multicolumn{1}{|c||}{Diagram} &
\multicolumn{1}{|c}{\# Moments} &
\multicolumn{1}{|c|}{Degree } &
\multicolumn{1}{|c||}{Order} & 
\multicolumn{1}{|c|}{\# Moments} &
\multicolumn{1}{|c|}{Degree } &
\multicolumn{1}{|c|}{Order} \\
\hline
$I_{1a}$ & 203& 26& 8& 15&  3&  2  \\
$I_{1b}$ & 269& 36& 9& 15&  3&  2  \\
$I_{2a}$ & 215& 31& 8& 19&  3&  3  \\
$I_{2b}$ & 269& 42& 9& 35&  6&  3  \\
$I_{4}$  & 623& 90&13& 131 & 24& 6 \\
\hline
\end{tabular}
\end{center}
\caption[]{Complexity of the smallest recurrences describing the integrals $I_{1a}$--$I_4$.}
\end{table}

\begin{table}[htbp]
  \begin{center}
  \renewcommand{\arraystretch}{1.3} 
  \begin{tabular}{|l||r|r|r||r|r|r||r|r|r||r|r|r|}
\hline
\multicolumn{1}{|c||}{ } &
\multicolumn{3}{|c||}{$\varepsilon^{-2}$} &
\multicolumn{3}{|c||}{$\varepsilon^{-1}$} &
\multicolumn{3}{|c||}{$\varepsilon^{0}$ rat.} &
\multicolumn{3}{|c|}{$\varepsilon^{0} \zeta_2$ } \\
\cline{2-13}
\multicolumn{1}{|c||}{Diagram} &
\multicolumn{1}{|c}{\# } &
\multicolumn{1}{|c|}{Deg. } &
\multicolumn{1}{|c||}{Ord.} & 
\multicolumn{1}{|c}{\# } &
\multicolumn{1}{|c|}{Deg. } &
\multicolumn{1}{|c||}{Ord.} & 
\multicolumn{1}{|c}{\# } &
\multicolumn{1}{|c|}{Deg. } &
\multicolumn{1}{|c||}{Ord.} & 
\multicolumn{1}{|c}{\# } &
\multicolumn{1}{|c|}{Deg. } &
\multicolumn{1}{|c|}{Ord.} \\
\hline
$I_{5a}$  &  15&  3& 2&  55& 11&  3& 142& 25& 5& 15& 3&  2  \\
$I_{5b}$  &  15&  3& 2&  55& 12&  3& 142& 27& 5& 15& 3&  2  \\
$I_{7a}$  &  19&  4& 2&  69& 14&  3& 164& 30& 5& 19& 4&  2  \\
$I_{7b}$  &  19&  4& 2&  79& 16&  3& 175& 34& 5& 19& 4&  2  \\
$I_{8}$   & 142& 26& 9& 463& 83& 10&1199&210&16&142&26&  5  \\
$I_{9a}$ &  47&  6& 4& 341& 57& 10& 949&156&16&109&17&  6  \\
$I_{9b}$ & 109& 17& 6& 323& 53& 10& 911&152&16& 47& 6&  4  \\
\hline
\end{tabular}
\end{center}
\caption[]{Complexity of the smallest recurrences describing the integrals $I_{5a}$--$I_{9b}$.}
\end{table}

The number of moments needed to determine the corresponding expressions ranges from $N 
= 142$ to 1199, except of the simpler pole- and $\zeta_3$-terms. On the other hand,
the most involved recurrence is of order 16 only, with larger polynomial coefficients 
of a degree up to 210. In Ref.~\cite{LINEQ} three of the present authors 
have handled far bigger recurrences of order 35 and degree $\sim 1000$. Still there is 
no thorough algorithm known producing the number of moments needed in case of the 
present integrals. 
\section{Conclusions}
\label{sec:5}
\renewcommand{\theequation}{\thesection.\arabic{equation}}
\setcounter{equation}{0} 

\vspace*{1mm}
\noindent
We calculated 3--loop Feynman integrals contributing to the ladder topologies of 
massive operator matrix elements up to six massive propagators with local operator
insertions on the massive quark line at general values of the Mellin variable $N$. 
The corresponding Feynman--parameter integrals are characterized by the Appell-function 
$F_1$ in case of six massive propagators. For lower numbers of massive propagators the
structures are simpler. The integrals can be turned into multiply nested sums. Most of them 
have been solved using modern summation technologies encoded in the package {\sf Sigma} 
in its present form automatically in product and difference fields. Here up to sixfold sums 
have been solved. This method applies also in case of integrals containing singularities in 
the dimensional parameter $\ep$ in the same way as for the constant term.

We also applied the method of hyperlogarithms in case of integrals which converge in $D=4$ 
dimensions, extended to the case of local operator insertions at general values of $N$. 
Up to now one of the diagrams could be solved by this method only. The polynomial expression
of the operator product insertion is first resummed using a continuous tracing parameter and
the integrals are then performed in terms of hyperlogarithms leading finally to iterated 
integrals over an alphabet of fixed numbers. It is then possible to symbolically convert 
the final expression into harmonic sums and their generalizations, with 
$\xi \in \{1,1/2,2\}$. 

All integrals are recurrent quantities in $N$ and their asymptotic representations are 
well behaved, since the growth $\sim 2^N$ present in individual contributions cancels. 
We also investigated the complexity of minimal difference equations associated with these 
integrals. The number of moments needed to reconstruct them directly varies between 
$\sim 200 ... 1200$, still being inaccessible by present methods. 

The present calculation shows that the topology class of ladder diagrams is widely 
accessible using the methods presented and can be applied in automated calculations. 
Generalized harmonic sums occur for individual diagrams. This, however, does not 
imply essential complications, since their analytic continuations and Mellin-inversions 
were derived in explicit form. Moreover, the corresponding functions are of a special 
type, which do not imply new numbers beyond the multiple zeta values in the limit $N 
\rightarrow \infty$. In various different steps of the present calculation the methods 
incorporated in the packages {\sf Sigma} \cite{SIGMA} and {\sf Harmonic Sums} \cite{HARMSU} 
were instrumental to obtain the corresponding solutions. Both packages have been updated 
and extended accordingly.

\vspace*{5mm}
\noindent
\paragraph{Acknowledgments.}~~We would like to thank I.~Bierenbaum, F.~Brown, A.~De Freitas, 
M.~Kauers, M.~Round, and V.~Stan for discussions. This work was supported in part by DFG 
Sonderforschungsbereich Transregio 9, Computergest\"utzte Theoretische Teilchenphysik, 
Studienstiftung des Deutschen Volkes, Austrian Science Fund (FWF) grant P203477-N18, 
the European Commission MRTN HEPTOOLS under Contract No. MRTN-CT-2006-035505, and ERC Starting Grant PAGAP 
FP7-257638.

\newpage
\begin{appendix}
\section{Feynman Rules}
\label{app:FR}

\vspace{1mm}
\noindent
In the present paper we have calculated scalar Feynman integrals with local operator 
insertions containing massive and massless scalar lines. The corresponding Feynman 
rules are somewhat different than in the case of QCD, cf. Sect.~8.1.~\cite{BBK2}, and 
read~: 
\begin{eqnarray}
&\text{\parbox[c]{.3\textwidth}{\includegraphics[scale=0.8]{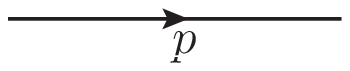}}}& \frac{1}{(p^2-m^2)}\\
&\text{\parbox[c]{.3\textwidth}{\includegraphics[scale=0.8]{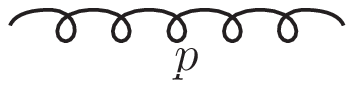}}}& \frac{1}{p^2} \\
&\text{\parbox[c]{.3\textwidth}{\includegraphics[scale=0.8]{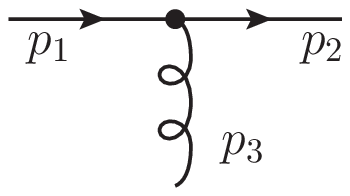}}}& g\\
&\text{\parbox[c]{.3\textwidth}{\includegraphics[scale=0.8]{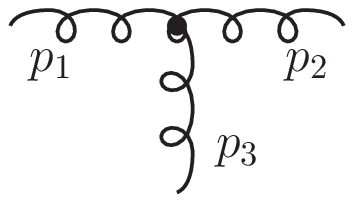}}}& g\\
&\text{\parbox[c]{.3\textwidth}{\includegraphics[scale=0.8]{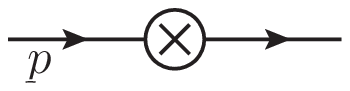}}}& (\Delta.p)^N\\
&\text{\parbox[c]{.3\textwidth}{\includegraphics[scale=0.8]{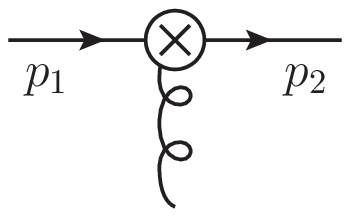}}}&
 g\sum_{j=0}^{N}(\Delta.p_1)^j(\Delta.p_2)^{N-j} \\
&\text{\parbox[c]{.3\textwidth}{\includegraphics[scale=0.8]{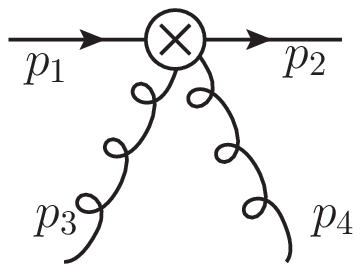}}}&
 \begin{array}{r}\displaystyle
 \text{a)} \quad g^2\sum_{j=0}^{N}\sum_{l=0}^{N-j}
          (\Delta.p_2)^j(\Delta.p_1)^{N-l-j} (\Delta.p_1+\Delta.p_4)^l\\
\displaystyle
 \text{b)} \quad g^2\sum_{j=0}^{N}\sum_{l=0}^{N-j}
          (\Delta.p_2)^j(\Delta.p_1)^{N-l-j} (\Delta.p_1+\Delta.p_3)^l
 \end{array}
\end{eqnarray}

\end{appendix}

\end{document}